\documentclass[twocolumn,prb,showpacs,preprintnumbers,amsmath,amssymb]{revtex4}

\usepackage{graphicx}
\usepackage{psfig}
\usepackage{dcolumn}
\usepackage{bm}

\begin{document}

\title{
Spin-flop transition in antiferromagnetic multilayers
}

\author{U.K. R\"o\ss ler}
\thanks
{Corresponding author 
}
\email{u.roessler@ifw-dresden.de}
\author{A.N.\ Bogdanov}
\altaffiliation[Permanent address: ]%
{Donetsk Institute for Physics and Technology,
340114 Donetsk, Ukraine
}
\email{a.bogdanov@ifw-dresden.de}
\affiliation{
Leibniz-Institut f{\"u}r Festk{\"o}rper- 
und Werkstoffforschung Dresden\\
Postfach 270116,
D--01171 Dresden, Germany
}%

\date{\today}

\begin{abstract}
{
A comprehensive theoretical investigation on 
the field-driven reorientation transitions 
in uniaxial multilayers with antiferromagnetic coupling is presented.
It is based on a complete survey of the
one-dimensional solutions for the basic phenomenological (micromagnetic) model 
that describes the magnetic properties of finite stacks 
made from ferromagnetic layers coupled 
antiferromagnetically through spacer layers.
The general structure of the phase diagrams is analysed.
At a high ratio of uniaxial anisotropy to antiferromagnetic interlayer exchange,
only a succession of collinear magnetic states is possible. 
With increasing field first-order (metamagnetic)
transitions occur from the antiferromagnetic ground-state
to a set of degenerate ferrimagnetic states
and to the saturated ferromagnetic state.
At low anisotropies, a first-order transition 
from the antiferromagnetic ground-state to 
an inhomogeneous spin-flop state occurs.
Between these two regions, transitional magnetic phases
occupy the range of intermediate anisotropies.
Detailed and quantitative phase diagrams are given for 
the basic model of antiferromagnetic multilayer systems with $N$ = 2 to 16 layers.
The connection of the phase diagrams 
with the spin-reorientation transitions 
in bulk antiferromagnets is discussed.
The limits of low anisotropy 
and large numbers of layers are 
analysed by two different representations of the magnetic energy,
namely, in terms of finite chains of staggered vectors 
and in a general continuum form.
It is shown that the phenomena widely described 
as ``surface spin-flop'' are driven only by the cut 
exchange interactions and the non-compensated magnetic
moment at the surface layers of a stacked antiferromagnetic system.
}
\end{abstract}


\maketitle

%
\section{Introduction}

Since the discovery of antiferromagnetic 
interlayer exchange \cite{Gruenberg87}
and the giant-magnetoresistance \cite{Baibich88}
in magnetic superlattices, 
such structures have become important components 
in magneto-electronic devices.
Research on these coupled multilayer systems is 
mainly driven by applications in magnetic storage 
technologies and the emerging spintronics.
\cite{Wolf01}
Specific structures with antiferromagnetic coupling 
are now considered as promising storage media.
\cite{Fullerton03}
It is clear, that applications 
necessitate a thorough control 
and understanding of their magnetic properties. 
On the other hand, such synthethic antiferromagnetic structures 
are ideal experimental models for studies of magnetic states 
and magnetization processes of antiferromagnets in confining geometries.
\cite{Wang94,Mills99,Hellwig03,Hellwig03b}
Two ferromagnetic layers coupled antiferromagnetically through a spacer, 
as the simplest of these systems, show properties 
which are formally described by the same phenomenological 
theory as a two-sublattice bulk antiferromagnet.
However, the magnetic states, 
domain structures, and magnetization processes 
even of such two-layer systems display a bewildering
variability and are far from understood in detail.%
\cite{Hubert98}
Modern experimental methods now allow 
imaging of magnetic states and domains 
in multilayer systems with 
resolution into the depths of multilayer stacks.
\cite{Felcher02,Lauter02}
Therefore, detailed studies of such 
structures have become feasible.

Theoretical models to describe the magnetic states of 
finite antiferromagnetic superlattices have revealed 
various surface effects, rich phase diagrams, 
and complex magnetization processes.
For antiferromagnetic layers, there are many other effects.
Surface-induced interactions, exchange couplings 
of antiferromagnets to other magnetic systems, 
in particular exchange bias in antiferromagnetic-ferromagnetic 
bilayer systems \cite{Nolting00} 
add to the multitude of possible magnetic states 
in antiferromagnetic layers.\cite{PRB03} 
However, the difficulty to probe and image magnetic structures 
in antiferromagnetic materials impedes the progress 
of our understanding on the antiferromagnetic side 
in such layered systems.
Hence, the finite antiferromagnetic superlattices 
are a suitably simple system which may promote a 
better understanding of surface related effects
in antiferromagnetic layered systems generally.
It is important to stress here, that surface effects 
in antiferromagnets have a different nature than 
in a ferromagnetic system. 
The cut exchange bonds at a (partially) uncompensated surface of 
an antiferromagnet causes a particular disbalance 
of magnetic forces which can never be understood
as a small surface-effect. 
In contrast, surface-effects
in ferromagnets are related to spin-orbit effects 
which usually are weak in comparison to the exchange.

A stack of magnetic layers with antiferromagnetic
couplings provides the basic model for 
cut exchange bonds at a fully uncompensated surface.
The study of the magnetic states and transitions for
such systems in external fields has a long history.
In 1968 Mills proposed that, 
at the surface of a uniaxial antiferromagnet,
a first-order transition should occur
in fields below the common ``bulk'' spin-flop (SF).
This transition from the antiferromagnetic 
to a ``surface spin-flop'' state should 
result in flopping a few layers of spins near the surface,
i.e. they would turn by nearly 90 degree.\cite{Mills68}
Further theoretical investigations
have improved the mathematical analysis of this
reorientation effect
\cite{Keffer73,Vernon78, Luthi83, Barron87, Lepage90}
however, direct experimental observations at
surfaces of crystalline
antiferromagnetic materials failed,
e.g., for the classical uniaxial antiferromagnet MnF$_2$,
(see bibliography and discussion in [\onlinecite{Wang94b}]).
\begin{figure}
\includegraphics[width=8.5cm]{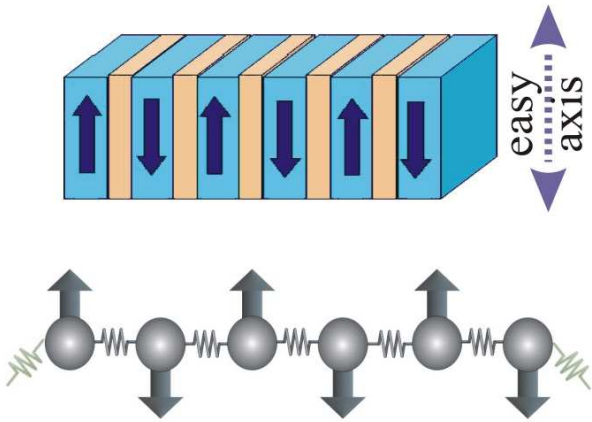}
\caption{%
\label{AFLayersCartoon}
(Color online)
Sketch of an antiferromagnetically coupled 
multilayer corresponding one-dimensional spin chain.
The "exchange springs" are cut at the ends of the finite chain.
} 
\end{figure}

As was mentioned above the basic model for 
two antiferromagnetically coupled layers 
is equivalent to the classical mean-field description of 
two-sublattice bulk antiferromagnets. \cite{Turov65,Stryjewski77}
These systems compose two large groups:
antiferromagnets with weak anisotropy \cite{Turov65} and 
strongly anisotropic uniaxial crystals that are commonly 
called \textit{metamagnets}.\cite{Stryjewski77}
The \textit{metamagnetic} phase transition 
between the antiferromagnetic and ferromagnetic
phases has been observed and investigated 
in many antiferromagnetic bulk systems.\cite{Stryjewski77}
Easy-axis antiferromagnets with
weak anisotropy  also compose
a large group of magnetically ordered crystals. 
For bulk antiferromagnets, the spin-flop transition has 
been predicted by N{\'e}el \cite{Neel36} 
and later was 
observed experimentally 
in CuCl${_2 \cdot}$ 2H ${_2}$O.\cite{Poulis52}
In the next fifty years, spin flop transitions have 
been discovered and carefully studied
in many classes of antiferromagnets
(see bibliography 
in Refs.~[\onlinecite{Turov65, UFN88, PRB02}]). 

The interest in spin-flops revived 
with the synthesis of magnetic multilayer stacks 
with indirect antiferromagnetic exchange coupling 
through spacers.\cite{Wang94}
These artificial antiferromagnetic layers,
with few magnetic units as macroscopic spins
instead of atomic spins (Fig.~\ref{AFLayersCartoon}) 
offer the possibility 
to study field-driven reorientation transitions
with unusually low exchange compared to anisotropies.\cite{Mills99}
Experimental investigations in 
Fe/Cr(211) antiferromagnetic superlattices
\cite{Wang94} seemed to confirm 
the scenario of the surface spin-flop transition 
introduced in Refs.~[\onlinecite{Mills68,Keffer73}].
The observation was supported by numerical investigation
of an elementary micromagnetic model.\cite{Wang94,Wang94b}
But, investigations on the magnetism 
of such antiferromagnetic superlattices and thin films
did not produce a consistent understanding of
the occurrence and nature of ``spin-flop'' transitions
or other reorientation transitions in layered
antiferromagnetic structures.
Recent experiments 
demonstrate complex behaviour
and different scenarios for the evolution of the magnetic states.
\cite{%
Howson94, Zabel94, Zabel96, Fullerton97,
Chesman98, Bab98, Hilt99, Temst00, Fitzsimmons00,
Nolting00,  Vavassori01, Felcher02, Lauter02, Nagy02, 
Aliev02, Lauter03}
The problem of the magnetic states in antiferromagnetic
superlattices with uniaxial anisotropy is strongly related 
to the long standing problem of a ``surface spin-flop''
discussed in various papers.
\cite{Vernon78, Luthi83, Barron87, Lepage90,
Wang94, Trallori94, Trallori95, Wang94b, Stamps94, Micheletti97,
Chung97, Rakhmanova98, Papan98, Momma98, Zhedanov98, Mills99, Dantas99,
Papan99, Papan00}
Theoretical investigations  are mostly based on
numerical calculations within a certain model, 
here called \textit{Mills model}.
\cite{Trallori94,Trallori95,Chung97,Rakhmanova98,Papan98,Momma98}
Further theoretical works related 
these finite or semi-infinite antiferromagnetic chain models 
to systems like 
the Frenkel-Kontorova model.\cite{Micheletti97,Trallori98}
These studies have led to controversial results
and generated a long-drawn discussion 
about the physical nature
of the reorientation transformations in the
antiferromagnetic superlattices. 
\cite{Trallori94,Chung97,Micheletti97,Rakhmanova98,
PRB03,PRB04}
Only few attempts have been made 
to gain a complete understanding
of the ground-state structure of 
the basic one-dimensional models 
for these superlattices.\cite{Micheletti97,Momma98,PRB04}
Thus, the basic questions, whether, when, and  how
a surface spin-flop occurs, were unresolved.

A full set of solutions for 
the model under discussion was
recently obtained by us.\cite{PRB04}
These outwardly simple systems
with few degrees of freedom own rich phase diagrams 
because of the competition between internal
stiffness and anisotropy in conjunctions with 
restricted dimensionality. 
Present paper presents an extended account 
and an analysis of 
the solutions from Refs.~[\onlinecite{PRB04,pss04}].
We explain the physical mechanism responsible for 
the formation of the main magnetic states 
in antiferromagnetic superlattices.
We derive a clear and simple
picture of the phenomena which have been discussed
as ``surface spin-flop''.
We demonstrate the connections with bulk antiferromagnetic
systems and other classes of magnetic nanostructures.
\cite{PRB04a,JMMM04}

The structure of the paper is as follows:
The  phenomenological model and its variants
are introduced in Sec.~\ref{SctModel}.
The analysis of possible magnetic states in the system
starts from two limiting cases of low and high uniaxial anisotropy.
The full solution for the generic highly symmetric {\em Mills model}
in applied fields along the easy axis are presented 
in Sec.~\ref{PhaseDiagram} and the general structure of 
the phase diagrams are presented. 
In this section, mainly analytical 
and numerical examples are employed 
to explain these solutions. 
Generalizations are briefly indicated. 
The methods can be extended to solve 
any other more general model for 
an antiferromagnetic multilayer system. 
In Sec.~\ref{WeakSection} we exploit the fact that 
the models for low-anisotropy antiferromagnetic superlattices 
can be reduced to equivalent models for a chain
of exchange coupled two-sublattice antiferromagnets.
This dimerization transformation reveals 
the physical mechanisms
ruling the magnetic states and reorientation transitions. 
Numerical results for the evolution of inhomogeneous spin-flop phases, 
magnetization curves, and phase diagrams are presented along with 
this discussion. 
In particular the limit of large numbers of layers 
and the emergence of the bulk spin-flop for such finite systems
are discussed.
Then, the continuum representation of the general models 
is presented. This offers a different point of view 
for the weak anisotropy. 
This continuum approximation 
is applicable to any weakly anisotropic system.
It allows to derive the structure of 
the inhomogeneous spin-flop phase 
for arbitrary models,
in particular model where the 
magnetic moments at the surfaces are partially
compensated. 
Hence, the succession of field-driven phase 
transitions between the antiferromagnetic state towards
the saturated ferromagnetic state via this spin-flop
phase can be completely analysed.
In Sec.~\ref{Puzzle} the general picture of magnetic states in
antiferromagnetic nanostructures and some recent 
experimental results are discussed
in the context of the new results.

\section{Model}\label{SctModel}

\subsection{The micromagnetic energy. Mills model}
\label{SbsMillsModel}
Let us consider a stack of $N$ ferromagnetic plates 
infinite in $x$- and $y$-directions 
and with finite thickness along the $z$-axis.
The magnetization of each plate is $\mathbf{m}_i$, 
and they are antiferromagnetically coupled through spacers.
Replacing this system by a chain of single-domain particles
with spontaneous magnetization $m_0^{(i)}= |{\mathbf m}_i|$, 
we may describe the magnetic configuration 
by the variables ${\mathbf s}_i={\mathbf m}_i/m_0^{(i)}$, 
i.e., by the set of unity vectors 
along the magnetization of the $i$th layer. 
We assume that the ferromagnetic layers have a 
uniaxial magnetic anisotropy with a common easy axis.
The phenomenological energy of this system 
can be written as
\begin{eqnarray}
{{\Theta}}_N=\sum_{i=1}^{N-1} \left[ J_i\,\mathbf{s}_i \cdot
\mathbf{s}_{i+1}
+ \widetilde{J}_i\,\left(\mathbf{s}_i \cdot
\mathbf{s}_{i+1} \right)^2 \right]
-\mathbf{H}\cdot \sum_{i=1}^{N}\,\zeta_i\, \mathbf{s}_i
\nonumber\\
- \frac{1}{2}\sum_{i=1}^{N} K_i\left( \mathbf{s}_i\cdot 
\mathbf{n} \right)^2
 - \sum_{i=1}^{N-1} K'_i\left( \mathbf{s}_i\cdot 
\mathbf{n} \right)
\left( \mathbf{s}_{i+1}\cdot \mathbf{n} \right) 
+ e_a (\mathbf{s}_i)\,.
\label{energy1}
\end{eqnarray}
Here, $\zeta_i = m_0^{(i)}/m_0$  designate
deviations of the magnetization in the $i$-th layer from the
average value $m_0$.
$J_i$ and $\tilde{J}_i$ are constants of bilinear 
and biquadratic exchange interactions, respectively.
The unity vector $\mathbf{n}$  points 
along the uniaxial anisotropy direction;
$K_i$ and $K'_i$ are constants of the in-plane and
inter-plane uniaxial anisotropy. 
Finally, $ e_a (\mathbf{m}_i)$
collects higher-order uniaxial and in-plane
magnetic anisotropy contributions,
e.g., intrinsic cubic (magnetocrystalline) anisotropy 
in systems like Fe and Ni layers.

The functional (\ref{energy1}) generalizes similar models
considered earlier in a number of studies
on magnetic states in antiferromagnetic 
multilayers with uniaxial anisotropy.
\cite{Wang94, Trallori94, Trallori95, Wang94b, Stamps94, Micheletti97,%
Rakhmanova98, Papan98, Momma98, Zhedanov98, Dantas99}
In a recent paper
on antiferromagnetic superlattice
with higher (tetragonal) symmetry, Ref.~[\onlinecite{PRB04a}], 
we have discussed
the general model and the justification of 
the approach used for this class of systems.
Therefore, the reader is referred to \cite{PRB04a} 
for a detailed discussion and further references.
Eq.~(\ref{energy1}) comprises the magnetic energies 
due to the main physical mechanisms,
which are present in magnetic multilayers
with indirect exchange 
through spacer layers (see Fig.~\ref{AFLayersCartoon}).
The ferromagnetic layers can be considered 
as homogeneously magnetized blocks with constant 
values of the magnetic interactions.
This assumption relies on the fact
that in ferromagnetic nanolayers
the intrinsic (direct) exchange coupling 
are usually very strong.
Thus, they play the dominating role 
for the magnetic order \textit{within} the layers
which react also very stiffly on all external 
and induced magnetic forces. 
Thus, the internal magnetic structure of an individual layer
experiences little change under influence of 
the induced magnetic forces at the surfaces and interfaces,
and the reorientation of other layers in the stack.
This hypothesis
has been justified by successful applications of such 
models to describe magnetization processes
in layered  ferro- and antiferromagnetic
nanostructures in different classes.
\cite{Johnson96, Thiaville92, PRL01,
Wang94, Zabel94, Ustinov01, Lauter02, PRB03}

Antiferromagnetic multilayers with uniaxial anisotropy 
and in applied field along the axis $\mathbf{n}$ show 
the strongest reorientation effects. 
Here, we address the overall magnetic properties 
of such uniaxial systems.
Their behaviour may be analysed by considering only terms with
bilinear exchange $J$, uniaxial intralayer anisotropy $K$ 
and an external field $H$.
The form of the energy in Eq.~(\ref{energy1}) considers 
additional terms, which are known to play a role
in antiferromagnetically coupled multilayers.
In particular, strong biquadratic exchange 
has been revealed in a number of antiferromagnetically 
coupled multilayers.\cite{Ruhrig91,Demokritov98, Chesman98, Ustinov01}
In Ref.~[\onlinecite{PRB04a}], we have studied 
the related phenomenological model for multilayers 
with zero and with four-fold (tetragonal) anisotropies
and we have discussed the relevance of 
biquadratic exchange ($\widetilde{J}\neq 0$),
which plays essentially only a quantitative role 
for the competition between the various possible 
magnetic states as long the antiferromagnetic 
ground-state remains collinear.

The uniaxial anisotropies may be 
intrinsic to the magnetic material of the film
or induced by surface effects. 
Thus, the uniaxial anisotropies
can be strongly changed with respect to bulk systems, 
and their strengths can be controlled 
in film system within wide margins.
Combination with intrinsic and induced fourth-order anisotropy, 
as considered in $e_a$ in Eq.~(\ref{energy1}),
may lead to peculiarities of magnetic properties,
see Refs.~[\onlinecite{Zabel94, PRB04a}], 
but here we will disregard these contributions.
Finally, strong demagnetization effects in antiferromagnetic
superlattices with perpendicular anisotropy are responsible
for complex evolution of multidomain states 
and specific magnetization processes.%
\cite{Hellwig03, Itoh03, Hellwig03b,JMMM04,Hellwig05} 
To investigate these effects 
the stray-field energy must be included in Eq.~(\ref{energy1}), 
and the corresponding magnetostatic problem has to be solved. 

The magnetic superlattices 
with antiferromagnetic coupling 
can be separated into two classes: 
\textit{non-compensated} systems with a net magnetization
and those with \textit{fully compensated} magnetization.
In the former case, the net magnetic moment 
strongly determines their global magnetic properties.
In many physical aspect their properties are similar
to those of bulk \textit{ferrimagnets}.
The main subject of this paper are the multilayers 
with fully compensated magnetization,i.e., 
multilayers with even $N$ and equal magnetization 
in all layers are similar to bulk collinear antiferromagnets. 
For simplicity, we assume $\zeta_i=1$
in all layers  $i=1\dots N$.
At the end of the paper we also consider effects
imposed by a partial compensation of the magnetization
in the endmost layers, i.e., deviations of $\zeta_1$
and $\zeta_N$ from unity.

For the reorientation effects in the antiferromagnetically 
coupled chain (Fig.~\ref{AFLayersCartoon}),
the effect due to the cut exchange bonds 
at the surface dominates.
The last moments in the chain are coupled 
only to one neighbour 
while ``internal'' moments interact with two. 
Thus, the moments at the surface 
experience a weakened exchange stiffness 
and are more susceptible to the reorienting 
influence of an applied field.
Due to this \textit{cut exchange} at the surfaces
the boundary moments oriented against the external field 
turn into the field direction 
in lower fields than internal moments. 
The simplified version of the model (\ref{energy1}) 
with equal constants $J_i=J$, $K_i=K$, 
and $\widetilde{J}_i = K'_i =e_a=0 $ for an applied field 
in direction of the easy axes, $\mathbf{H} \parallel \mathbf{n}$,
describes  the effect of the cut bonds 
as the sole surface-imposed factor. 
It allows to investigate this surface effect
separately from other interactions.

Usually the magnetization of the layers 
is confined to a certain plane. 
For many multilayers, this is the film plane 
owing to demagnetization.
For this case the deviations of $\mathbf{m}_i$
from the anisotropy axis can be described by 
angles $\theta_i$, and the energy (\ref{energy1})
in the equal constant model
is reduced to the following form 
\begin{eqnarray}
\Phi_N &   =  & 
J\sum_{i=1}^{N-1} \cos(\theta_i - \theta_{i+1}) \\
 & & -H \sum_{i=1}^{N}\cos\theta_i
- \frac{K}{2}\sum_{i=1}^{N} \cos^2 \theta_i\,. \nonumber
\label{energyMills}
\end{eqnarray}
For $J > 0$ and $K > 0$, the Eq.~(\ref{energyMills})
describes antiferromagnetically coupled 
ferromagnetic layers with easy-axis anisotropy 
in an external field along the anisotropy axis.
Energy~(\ref{energyMills})
has been introduced by Mills
for a semi-infinite chain ($N= \infty$).\cite{Mills68}
Later this model (called here \textit{Mills model})
has been intensively studied 
for finite and infinite $N$
\cite{Keffer73, Vernon78, Luthi83, Barron87, Lepage90,
Wang94, Trallori94, Trallori95, Wang94b, Stamps94, Micheletti97,
Chung97, Rakhmanova98, Papan98, Momma98, Zhedanov98, Mills99, Dantas99,
Papan99, Papan00}
and has been used as basic ansatz
to analyze experimental results 
in antiferromagnetically 
coupled multilayers.
\cite{Wang94, Felcher02, Lauter03}

Both the exchange interactions and uniaxial anisotropy
in (\ref{energyMills}) have surface/interface induced nature, 
their values are very sensitive to many physical factors 
such as the structure of the spacers and substrates.
\cite{Johnson96, Moriarty01}
Hence, the materials parameters 
may vary from layer to layer in the stack.
A generalization of the Mills model
may include differing parameters for each layer 
%
%
\begin{eqnarray}
\widehat{\Phi}_N &   =  & 
\sum_{i=1}^{N-1} J_i\, \cos(\theta_i - \theta_{i+1}) \\
& & -H \sum_{i=1}^{N}\zeta_i\,\cos(\theta_i)
- \frac{1}{2}\sum_{i=1}^{N} K_i \cos^2 \theta_i\,. \nonumber
\label{energy2}
\end{eqnarray}
This is a general model for antiferromagnetic multilayers.
Even for a stack of {\em identical} nanolayers,
the top and bottom layers
still have a different ``neighbourhood''
than internal layers.
To describe these effects we may 
introduce a \textit{modified Mills model}
with equal materials parameters 
for all internal layers
$J_i = J$ ($i = 2, 3,$\,\dots\,$, N-2$), 
$K_i = K$ ($i = 2, 3, $\,\dots\,$, N-1$) and 
different values for the first and last
layer ($J_1=J_{N-1}=J_s$, $K_1=K_N = K_s$,
$\zeta_1=\zeta_N=\zeta_s < 1$)
\begin{eqnarray}
\widetilde{\Phi}_N  &=& J_s \left[\cos(\theta_1 - \theta_{2})+
\cos(\theta_{N-1} - \theta_{N}\right] \\
 & & + J\sum_{i=2}^{N-2}  \cos(\theta_i - \theta_{i+1}) \nonumber \\
& & -H \zeta_s \left(\cos \theta_1+\cos \theta_N \right)
 -H \sum_{i=2}^{N-1}\cos \theta_i \nonumber\\
& & - \frac{1}{2}K_s \left(\cos^2 \theta_1+ \cos^2 \theta_N \right)
 - \frac{K}{2}\sum_{i=2}^{N-1} \cos^2 \theta_i\,. \nonumber
\label{MillsModified}
\end{eqnarray}
In all these cases, calculations of 
the magnetic states for 
the antiferromagnetic superlattices 
can be reduced to the minimization  
of the energy functions $\Phi_N (\theta_1, \theta_2,\,\dots\,,\theta_N)$.
In this paper, we study in detail solutions 
for the chains with equal constants 
as described by \textit{Mills} model (\ref{energyMills}),
and we discuss the generalizations according to
Eqs.~(\ref{energy2}) and (\ref{MillsModified}).

\subsection{General features of the solutions. 
Relation to bulk antiferromagnetism}

The antiferromagnetic multilayers 
with $N=2$ 
are of particular importance for 
investigations 
on surface/interface-induced interactions.
\cite{Zabel94, Chesman98, Bab98, Vavassori01}
In experimental works they are 
often referred as ``trilayers'',
we use here the term ``two-layer systems''.
The energy (\ref{energyMills}) 
for $N=2$ is the same function as the 
mean-field magnetic energy 
of a \textit{bulk two-sublattice} antiferromagnet
\begin{eqnarray}
\widetilde{\Phi}_2  & = & J\, \mathbf{s}_1 \cdot \mathbf{s}_2
-H\, [\left( \mathbf{s}_1\cdot \mathbf{n} \right) 
+\left(\mathbf{s}_2\cdot \mathbf{n} \right)] \nonumber \\
& & - \frac{K}{2}\left[
\left( \mathbf{s}_1\cdot \mathbf{n} \right)^2  + 
\left( \mathbf{s}_2\cdot \mathbf{n} \right)^2  
\right]\,.
\label{energy2a}
\end{eqnarray}

\begin{figure}
\includegraphics[width=8.5cm]{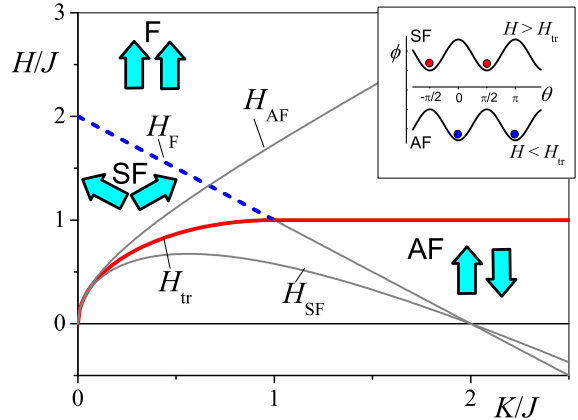}
\caption{
\label{PhDiagN2}
(Color online)
The phase diagram of the solutions for a two-layer system
includes antiferromagnetic (AF), spin-flop(SF) 
and ferromagnetic (F) phases.
The AF phase is separated from the SF and F phases
by the first-order transition lines (thick). 
The second-order ``spin-flip'' line $H_{\mathrm{F}}$ (dashed)
separates SF and  and F phases. 
Thin lines indicate 
stability limits of the corresponding phases.
Inset: in the spin-flop field $H = H_{\mathrm{tr}}$ the potential
wells switch from ($0$, $\pi$) (AF phase) to
($-\pi/2$, $\pi/2$) (SF phase).
}
\end{figure}

We briefly review the reorientation transitions
in these two-layers systems to fix our notation and terminology.
Following a general convention
we introduce the linear combinations of the magnetization vectors
$\mathbf{s}_{1,2} = \mathbf{M} \pm  \mathbf{L}$,
the \textit{total} or \textit{net  magnetization} $\mathbf{M}$
and the \textit{staggered} magnetization $\mathbf{L}$,
that is also called vector of \textit{antiferromagnetic order} 
\cite{Turov65}. 
The equations $|\mathbf{s}_i|  = 1$ lead to the
constraints $\mathbf{M}^2+ \mathbf{L}^2 = 1$ and
$(\mathbf{M}\cdot\mathbf{L}) = 0$.
In a magnetic field along the easy axis
the vectors $\mathbf{H}$, $\mathbf{s}_1$ and 
$\mathbf{s}_2$ lie in a fixed plane. 
The magnitude of the net magnetization $M =| \mathbf{M}|$
can be used as an internal parameter.
After an independent minimization with respect to $M$,
the energy and the net magnetization depend only 
on the orientation of the staggered vector $\mathbf{L}$
and can be expressed as functions of the angle $\phi$ 
between anisotropy axis $\mathbf{n}$ and $\mathbf{L}$ 
\begin{eqnarray}
\widetilde{\Phi}_2 (\phi) &  = & -\frac{H^2 \sin^2 \phi}{(2J+K \cos 2 \phi) }
-K \cos^2 \phi +J, \\
M &  = &  \frac{2H \sin \phi}{ (2J+K \cos 2 \phi)}\,.
\label{energy2b}
\end{eqnarray}
Then, the phase diagram of the antiferromagnetic two-layer system
is given by the potential energy 
for a system with one variable $\phi$ (\ref{energy2b}) 
with the control of materials parameters $K/J, H/J$.
This phase diagram is plotted in Fig.~\ref{PhDiagN2}.
The structure of the phase diagram is determined
by the following characteristic fields
\begin{eqnarray}
H_{\mathrm{tr}} = \sqrt{K(2J-K)}, \; 
H_{\mathrm{FM}} = J,\; H_{\mathrm{F}}=(2J-K)
\nonumber\\
H_{\mathrm{AF}} = \sqrt{K(2J+K)},\;
H_{\mathrm{SF}} = (2J-K) \sqrt{\frac{K}{2J+K}}\,. 
\label{critfields2b}
\end{eqnarray}
For $K < J$ a first-order transition
between the antiferromagnetic (AF) phase with $\phi = 0, \pi$
and the spin-flop phase (SF) with $\phi = \pm \pi/2$
occurs at the \textit{spin-flop} field $H_{\mathrm{tr}}$.
The fields $H_{\mathrm{AF}}$, $H_{\mathrm{SF}}$ are stability limits of these
two competing phases. The difference $H_{\mathrm{AF}}- H_{\mathrm{SF}}$ gives 
the width of the metastability region.
Because the antiferromagnetic state
has zero magnetization 
the magnetization jump at the first-order transition, 
$\Delta M = m_0\, M_{\mathrm{SF}}(H_{\mathrm{tr}}) = m_0\,\sqrt{K/(2J-K)}$ 
exactly in the field $H_{\mathrm{tr}}$,
equals the magnetization of the spin-flop phase.
For increasing anisotropy $K$ this magnetization jump
$M_{\mathrm{SF}}(H_{\mathrm{SF}})$ gradually increases
from very small values $ M \ll 1$  for $K \ll J$ 
to the saturation value  $M = 1$ which marks the 
end point of the first-order transition line 
between the AF and SF phase that is reached at $K=J$.
A continuous transition from the
SF phase to the ferromagnetic (F) phase 
occurs at $H_{\mathrm{F}}$. 
This transition leading to an enforced field-polarized state 
is usually referred to as \textit{spin-flip} transition.

For $K>J$, the SF phase does not arise as a stable state;
instead there is a direct first-order transition
between the AF and F phase at $H_1 =J$.
Such transitions in antiferromagnets are known 
as \textit{metamagnetic} transition.\cite{Stryjewski77}
For this high anisotropy region, $K > J$, 
the critical field $H_{\mathrm{F}}$ plays the role of 
the stability limit for the ferromagnetic phase (Fig. \ref{PhDiagN2}).
The metamagnetic transition is characterized 
by a large jump of the magnetization 
$ \Delta M (H_1) = m_0$ 
and extremely broad metastability regions.

For low-anisotropy systems in the region $K \ll J$,
the energy (\ref{energy2b}) can be simplified 
\begin{eqnarray}
\widetilde{\Phi}_2 (\phi) = 
\left(\frac{H^2 -H_0^2}{4J}\right) \cos 2 \phi , \qquad H_0 = \sqrt{2JK} 
\,.
\label{energy2c}
\end{eqnarray} 
In this limit the metastable region is restricted to a close vicinity
of the spin-flop field:
$H_{\mathrm{AF}}\approx H_{\mathrm{SF}}\approx H_{\mathrm{tr}}\approx H_0$.
This simplified potential energy Eq.~(\ref{energy2c}) 
reveals the physical mechanism of the spin-flop transition.
At zero field the uniaxial anisotropy 
stabilizes the antiferromagnetic phase.
The potential wells at $\phi =0, \pi$ 
corresponding to the two antiferromagnetic 
states are stable.
An increasing applied field, $H < H_{\mathrm{tr}}$,
gradually  reduces the height of the
potential barrier between 
the antiferromagnetic states. 
At the threshold spin-flop field $H_{\mathrm{tr}}$
the stable potential wells switch 
into the configurations $\phi =-\pi/2$, $\phi =\pi/2$
that correspond to the flopped states (Fig. \ref{PhDiagN2}).

Contrary to natural antiferromagnetic crystals
which are described only by marginal parts of the $K/J$ scale
with either low anisotropy, $K/J\ll 1$, or very large anisotropy, $K\gg 1$, 
antiferromagnets composed of magnetic nanolayers 
can have arbitrary values of $K/J$. 
These artificial antiferromagnets cover 
the whole phase plane in Fig. \ref{PhDiagN2}.

\begin{figure}
\centerline{\includegraphics[width=8.5cm]{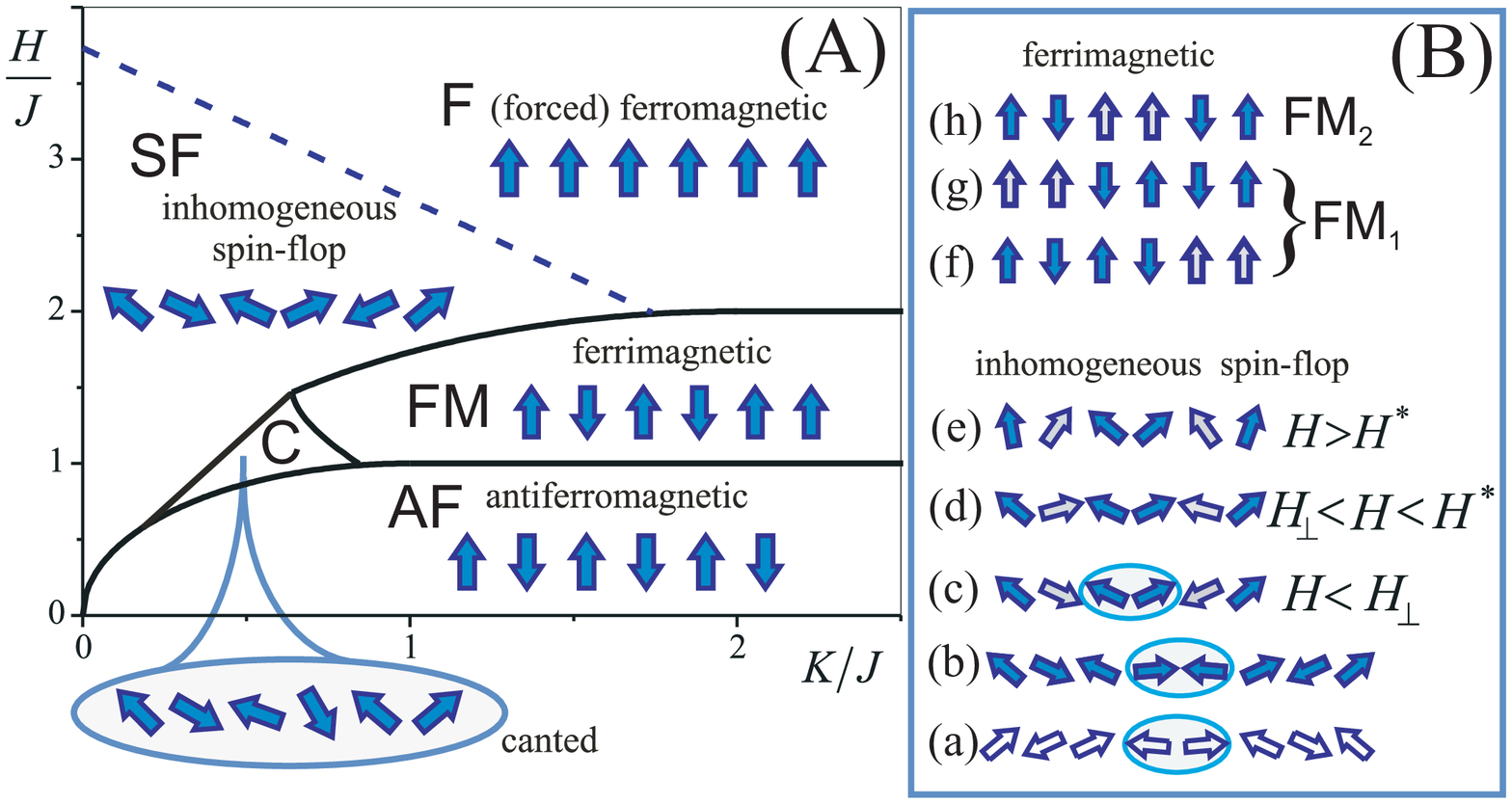}}
\caption{
\label{Phases}
(Color online)
Sketch of the phase diagram
for $N = 6$ introduces the main types
of the solutions for Mills model Eq.~(\ref{energyMills})
with arbitrary even $N$ (A).
Due to the cut exchange bonds 
the flopped states are spatially inhomogeneous
and can exist as symmetric \textit{inhomogeneous spin-flop} phase (SF)
for low anisotropy, or 
as asymmetric \textit{canted} phases (C) 
for moderate anisotropies.
In strongly anisotropic systems ($K \geq J$)
spin configurations with flipped spins exist
These collinear \textit{ferrimagnetic} (FM) phases exist at
intermediate fields
between the \textit{antiferromagnetic} 
and \textit{ferromagnetic} states.
The spin configurations in panel (B) 
show for $N=8$ the degeneracy of 
the inhomogeneous SF state for $N=8$ (a) and (b); 
for $N=6$ the evolution of the inhomogeneous SF 
in increasing field (c) to (e) 
and the two different types of FM states, 
where FM$_1$ is degenerate (f), (h) and FM$_2$ (h) 
is a collinear version of the inhomogeneous 
SF state (c)-(e) in an even-odd system.
}
\end{figure}

In finite multilayers with $N > 2$
the cutting of the exchange bonds 
at the surfaces (Fig. \ref{AFLayersCartoon})
causes a strong disbalance of the 
exchange interactions along the chain.
This disbalance is the determining factor 
for the appearance of magnetic states in the system.
The detailed analysis of the solutions for
Mills model will be given in the next section.
Here, we summarize the physical mechanism 
ruling the formation of magnetic configurations 
in simple terms.
In the antiferromagnetic configuration 
the moments at one surface always point against an applied field. 
These moments can be reversed 
more easily than the internal moments 
because of the cut exchange.
Depending on the relative strength of 
the exchange and anisotropy,
this specific instability leads 
to different reoriented configurations.
At very strong uniaxial anisotropy ($K \gg J$) 
the exchange coupling between layers becomes negligible.
In this case the reorientation of the magnetization
in the endmost layer does not influence
magnetic states in the other layers.
In an increasing field, a collinear spin configuration 
with an inverted endmost moment is reached.
The corresponding \textit{ferrimagnetic} (FM) phase
becomes energetically stable at a field $H_{\mathrm{FM}} = J$ 
through a discontinuous first-order transition
(Fig. \ref{Phases}).
\cite{PRB04,JMMM04}
In the opposite case of weak anisotropy, $K \ll J$,
the exchange coupling plays the dominating role.
Accordingly the overturn of the endmost moment
is spread over the entire stack and creates a
spatially inhomogeneous spin configuration
(Fig. \ref{Phases}).
\cite{PRB04,PRB04a}
In Ref.~[\onlinecite{PRB04}] this mode was called
\textit{inhomogeneous spin-flop phase}.
In increasing fields,
a curious evolution 
takes place within this spin-flop phase,
where some moments rotate against 
the applied field and change their sense of rotation 
at higher fields.\cite{PRB04a} 
A continuous spin-flip into the saturated state occurs 
at an ``exchange'' field $H_{\mathrm{E}}$, 
which depends on  the number of layers 
and on the anisotropy $K$.
In the region of moderate anisotropy
spatially inhomogeneous asymmetric states
exist as transitional phases between
the inhomogeneous SF and FM phases.
These asymmetric phases arise 
by canting transitions, i.e., 
elastic distortions
of the collinear FM phases when $K/J \stackrel{<}{{\sim}} 1$.
These asymmetric \textit{canted} (C) phases 
can be considered as superpositions
of ferrimagnetic states and the inhomogeneous spin-flop state
(Fig.~\ref{Phases}). 
This means that in these low symmetry intermediate C phases
all the symmetries are broken that are broken in the 
corresponding SF and FM phases.
Thus, magnetic states arising in Mills model 
comprise antiferromagnetic, spin-flop, and ferromagnetic phases, 
which exist in bulk antiferromagnets,
and additional ferrimagnetic and canted configurations. 
The latter phases are imposed by the exchange cut.
They are specific to finite antiferromagnetic layer systems.
In particular, the detailed solutions for 
larger $N$ show series of different canted phases.
The corresponding phase diagrams 
include a large number of critical points 
and a tangled net of transition and lability lines
(see examples in [\onlinecite{Momma98,PRB04,pss04}]).
The cut exchange bonds underly this complexity 
as the general physical mechanism for the formation 
of the various inhomogeneous magnetic states and their
transitions into the simple collinear states in the
limiting regions of the phase diagram 
for and low high anisotropy and for large fields.
Therefore, all these phase diagrams 
have a general topology represented in Fig.~\ref{Phases}. 

\section{Phase diagram of the solutions}\label{PhaseDiagram}

In this section we analyse the solutions
for Mills model (\ref{energyMills}), derive the regions
of their existence, and conditions of 
the transitions between different phases.

\begin{figure}
\includegraphics[width=8.5cm]{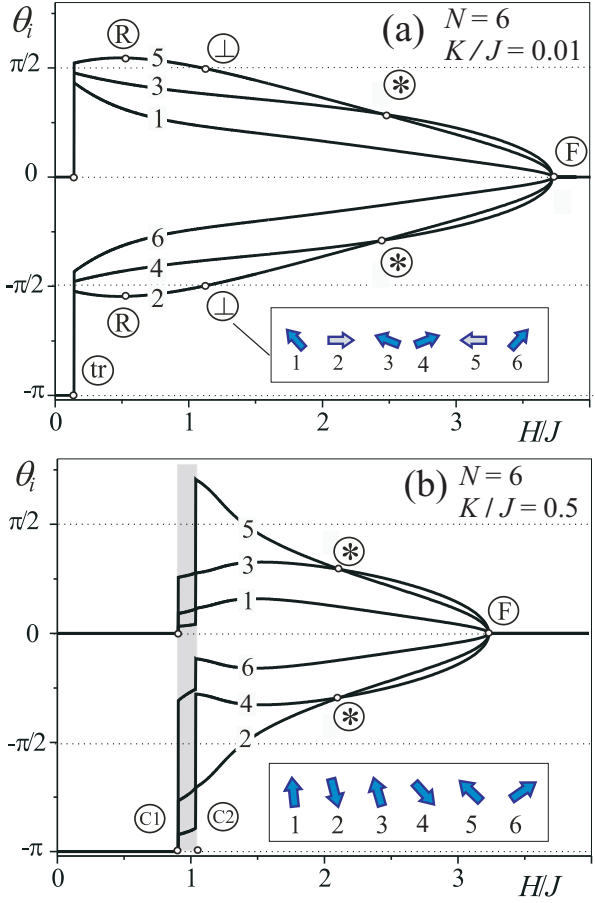} 
\caption{
\label{N6Profiles}
(Color online)
Evolution of the equilibrium magnetic configurations $\theta_i$ 
in Mills model for antiferromagnetic $N=6$ layers 
with finite uniaxial anisotropy, $K = 0.01\,J$ (a), $K = 0.5\,J$ (b),
in magnetic fields applied in direction of the easy axis.
For the low anisotropy case only AF, SF, and F states exist (a).
The points indicate the spin-flop field $H_{\mathrm{tr}}$ and the spin flip
field $H_{\mathrm{F}}$. 
Characteristic fields are $H_{\mathrm{R}}$, $H_{\bot}$, and  $H^*$
as marked in the plot.
The spin configurations correspond to
$H = H_{\bot}= 1.104\,J$ where the
magnetization in the 2th and 5th
layer is perpendicular to the applied field.
For intermediate anisotropy (b)  
canted asymmetric states exist in
the field interval $H_{\mathrm{C1}} < H < H_{\mathrm{C2}}$.
The spin configuration in (b) corresponds to
the canted phase at the transition field
$H_{\mathrm{C1}} = 0.906\,J$.
}
\end{figure}

\subsection{Spin-flop transition and solutions 
for inhomogeneous spin-flop phases}

We start with low-anisotropy systems ($J \gg K$).
Here, we consider generalized models 
that keep mirror symmetry about the multilayer center
with parameters $J_i = J_{N-i}$, $K_i = K_{N+1-i}$, etc.,
for $i=1\dots N-2$ or $N-1$, respectively.
Then, the equations, that minimize the energy of these systems,
have solutions for an inhomogeneous spin-flop phase 
with the property $\theta_{i} = - \theta_{N-i+1}$.\cite{PRB04,PRB04a}
These magnetic configurations have different 
properties when $N=4\,l$, i.e., $N$ is divisible by four
called here \textit{even-even} systems, 
or when $N=4\,l+2$ for \textit{even-odd} systems ($l= 1,2...$). 

In low fields, ${J_i \gg  H}$, the spins
in the flopped state deviate only slightly 
from the the direction perpendicular to the 
easy axis
\begin{eqnarray}
\label{SFsol}
\theta_{2j-1} = \pi/2 - \alpha_{2j-1}, \;
\theta_{2j} = -\pi/2 + \alpha_{N-2j+1},  \\
|\alpha_{2j-1}| \ll 1 \qquad j=1,\dots N/2\,. \nonumber
\end{eqnarray}
The expansion of energy (\ref{energy1}) with respect
to the small parameters $\alpha_{2j-1}$ allows to derive
analytical solutions for the flopped states that can 
be formulated for arbitrary models.
As an illustration we write the parameters $\alpha_{2j-1}$
for Mills model (\ref{energyMills}), i.e., with equal constants 
in the even-even case $N= 8$ 
\begin{eqnarray}
\label{sol8N}
\alpha_1 = \frac{2H}{J} \left( 1 + \frac{11}{2} \frac{K}{J} \right), \;
\alpha_3 = \frac{H}{J} \left( 1 + \frac{6K}{J} \right), \\
\alpha_{5} = \frac{2H}{J}  \frac{K}{J}, \;
\alpha_{7} = \frac{H}{J} \left( 1 +  \frac{9K}{J} \right), \nonumber
\end{eqnarray}
and in the even-odd case $N = 10$ 
\begin{eqnarray}
\label{sol10N}
& & \alpha_1 =  \frac{5}{2}\frac{H}{J} 
\left( 1 + \frac{17}{2} \frac{K}{J} \right), \;
\alpha_3 = \frac{3}{2} \frac{H}{J} 
\left( 1 + \frac{59}{6} \frac{K}{J} \right), \nonumber\\
& & \alpha_5 = \frac{1}{2} \frac{H}{J} 
\left( 1 + \frac{13}{2} \frac{K}{J} \right), 
\alpha_{7} =  \frac{1}{2} \frac{H}{J} 
\left( 1 + \frac{37}{2} \frac{K}{J} \right), \nonumber\\
& &\alpha_{9} = \frac{3}{2}\frac{H}{J} 
\left( 1 + \frac{25}{2} \frac{K}{J} \right)\,. 
\end{eqnarray}

The Eqs.~(\ref{SFsol}), (\ref{sol8N}), and (\ref{sol10N}) 
display the generic structure of these solutions which 
apply also for generalized models with arbitrary parameters. 
The deviations of the magnetization direction from the directions $\pm\pi/2$ 
are small of the order $H/J$.
The corrections due to the anisotropy are of the order $(HK)/J^2$. 
The deviations of the magnetization direction in the different layers 
gradually increase towards the endmost layers $i = 1$ or $i = N$, respectively.
The dominating exchange interactions 
favour antiparallel ordering in the adjacent layers.
In the flopped configurations (\ref{sol8N}), (\ref{sol10N}) 
pairs of spins essentially remain antiparallel.
E.g., for $N=8$, the interior pairs $\mathbf{s}_{2}$, $\mathbf{s}_{3}$ and
$\mathbf{s}_{6}$, $\mathbf{s}_{7}$ are almost antiparallel 
(Fig. \ref{Phases} B, panel (a)). 
This fact may be stated more precisely. 
According to (\ref{sol8N}) 
$|\theta_3 - \theta_2| = |\theta_7 - \theta_6| =
\pi - (3HK)/J^2$, this means the slight deviations 
from antiparallel arrangement are due to a second order effect.
The exchange coupling in such pairs is stronger than
the Zeeman energy of the pair. 
This causes an interesting effect,
a \textit{reverse} rotation of the magnetization in
a number of layers.
In the multilayer with $N=8$, the magnetizations $\mathbf{s}_{2}$, $\mathbf{s}_{7}$ 
undergo such a reverse rotation according to Eqs.~(\ref{sol8N}); 
for the case $N=10$ the corresponding magnetizations are
$\mathbf{s}_{2}$, $\mathbf{s}_{4}$, $\mathbf{s}_{7}$, and $\mathbf{s}_{9}$,
see Eqs.~(\ref{sol10N}).

For even-odd systems, the projections of the magnetization vectors for the central
layers ($\mathbf{s}_{N/2}$, $\mathbf{s}_{N/2+1}$) 
onto the field direction 
are of the order $~H/J$, which
is much larger than the corresponding projections of the order $~(HK)/J^2$
in even-even systems 
(compare solutions for $\theta_4, \theta_5$ in Eqs.~(\ref{sol8N}), 
$\theta_5, \theta_6$ (\ref{sol10N}) and spin configurations
in panels (a) and (c) Inset B of Fig.~\ref{Phases}, respectively).
The solutions $\theta_i (H)$ for Mills model
of a multilayer with $N = 6$ in Fig. \ref{N6Profiles}
illustrate the general features of the field induced
evolution of the spin-flop state.
(See also the configurations in panels (c)-(e) in Inset B 
of  Fig.~\ref{Phases},
and the solutions for $N = 16$ in Ref.~[\onlinecite{PRB04}] and
$N = 12$ in Ref.~[\onlinecite{pss04}]).
An increasing magnetic field gradually
slows down the reverse rotation of the spins
with negative projections onto the field direction.
Finally at characteristic fields $H_R^{(i)}$ 
the sense of rotation changes. 
In this point $d \theta_i/dH=0$.
Another set of characteristic fields $H_{\bot}^{(i)}$ defines
the points where the projection of $\mathbf{s}_i$ changes the sign, i.e., 
$(\mathbf{s}_i(H_{\bot}^{(i)})\cdot \mathbf{H}) = 0$.
In an increasing field these characteristic fields, $H_R^{(i)}$ and then
$H_{\bot}^{(i)}$, are reached first for the central layers and at higher fields 
for those closer to the boundaries.
Fig.~\ref{N6Profiles} (a) displays
angles $\theta_i$ for 
an example of Mills model Eq.~(\ref{energyMills}).
with low $K\neq0$.
Finally, for Mills model at a particular field $H^{\star}$ 
the projection of the magnetization onto the field direction 
is equal for all interior layers. 
This knot point is designated by ``$\star$''. 
The value of $H^{\star}$ is analytically given by 
\begin{eqnarray}
H^{\star}& =&  2\sqrt{3}J \left(1-\frac{K}{4J}\right)
\left(1+\frac{K}{3J}\right)^{1/2}\times \\
& & \times \left[1+\frac{3K}{4J}+
\sqrt{1+\left(\frac{K}{4J}\right)^2\,}\;\right]^{-1/2}\,. \nonumber
\label{Hstar}
\end{eqnarray}
In particular, for $K = 0$ the knot point is $H^{\star}=\sqrt{6}\,H_{\mathrm{F}}/4$,
which coincides with the value derived in Ref.~[\onlinecite{PRB04a}].
For $H > H^{\star}$ the positive projections of the magnetization
onto the direction of the magnetic field decreases towards the center.
The inhomogeneous spin-flop states for
isotropic Mills models, $K\equiv 0$, exist 
starting from zero field and have similar features 
as those for finite anisotropy.
\cite{PRB04a}

\subsection{Critical lines and multicritical points}

\begin{figure*}
\includegraphics[width=8.5cm]{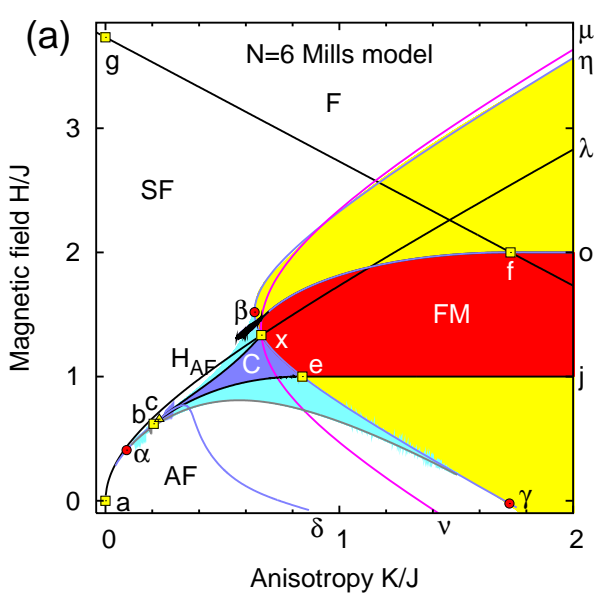}
\includegraphics[width=8.5cm]{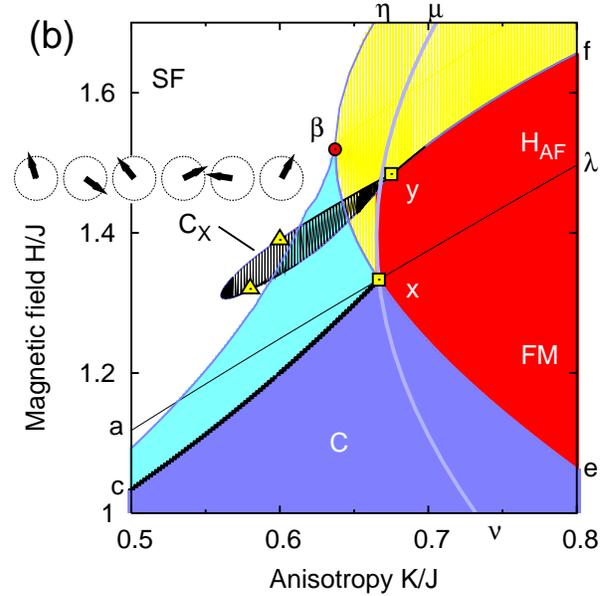}
\caption{
\label{PDN6}
(Color online)
Phase diagram for Mills model with $N$=6 (a).
Detail of phase diagram for Mills model 
with $N$=6 (b).
Special canted states C$_\mathrm{X}$ 
(sketched configuration)
are stable in the hatched area. 
The point $y$ is a triple point, 
where FM, SF, and C$_\mathrm{X}$ phase coexist.
Triangles designate tricritical points,
where the SF-C$_\mathrm{X}$ transitions
change from continuous to first-order 
with increasing anisotropy. 
}
\end{figure*}
In this section we determine stability regions
of the solutions, conditions of the phase
transitions between them and construct the
phase diagrams.
Mills model (\ref{energyMills}) has three independent
control parameters $K/J$, $H/J$, and $N$.
Correspondingly a set of ($K/J,\, H/J$) phase planes
for different $N$ provides complete information about
the solutions for the model (\ref{energyMills}).  
As illustration we  present the ($K/J,\, H/J$) phase diagram 
for Mills model with $N = 6$ (Fig.~\ref{PDN6}), which 
containts all essential features of the generic phase diagram
and some complications (Fig.~\ref{PDN6}(b)), which are absent 
in the simplest case $N=4$ as presented in Ref.~[\onlinecite{PRB04}].
The essential critical points for Mills models with $N = 4 \dots 16$ 
are given in Table I. 
Then we proceed to discuss general features of the model with arbitrary $N$.

To determine the conditions for the phase transitions 
between different spin configurations
and the stability regions of these phases we may use standard
procedures.
The equality of the equilibrium energies of the competing
phases  yields the condition of the first-order transitions.
The stability of the solutions 
$\{\theta_i\}_{i=1,\dots,N}$  
can be checked 
by writing the energy of the system for small arbitrary distortions,
$\tilde{\theta}_i = \theta_i +\delta \theta_i$, which yields
the expansion 
\begin{eqnarray}
 \Phi_N(\tilde{\theta}_i) & = &  \Phi_N(\theta_i)
+ \sum_{i,j=1}^{N}A_{ij}\delta \theta_i\delta \theta_j, \\
& &A_{ij}= \partial^2\Phi_N/\partial \theta_i \partial \theta_j\,. \nonumber
\label{distortedEnergy}
\end{eqnarray}
The solutions  $\theta_i$ are stable, 
if all eigenvalue of the symmetric matrix 
$\mathbf{A}_N =(( A_{ij}))$ are positive. 
In particular, for the AF phase within Mills model
the matrix $\mathbf{A}_N$ has a band structure given by
$A_{2j-1,\, 2j-1}=J+K+H$, $A_{2j,\, 2j}=J+K-H$ for $j=1 \dots N/2$, and
$A_{i,\, i+1}= A_{i+1,\, i}= J$ for $j=1 \dots N-1$. 
All other elements are equal zero. 
The determinant $D_N=\det(\mathbf{A}_N)$ 
can be reduced to the following form
\begin{eqnarray}
D_N  & = & 
\underbrace{[H^2-H_{\mathrm{AF}}^2]}_{D_2} \times \\
 & & \times \{[H^2-(K^2+4JK+2J^2)]D_{N-4}  \nonumber\\
   & & +[H^2-(K^2+4JK+3J^2)]\sum_{i=3}^{N/2}(-1)^{i}D_{N-2i}\}\,.\nonumber
\label{matrixAF2}
\end{eqnarray}
The obvious convention $D_0  = 1$ and $D_k= 0$ for $k < 0$
starts the recursion in Eq.~(\ref{matrixAF2}).
Any determinant $D_N$ includes the determinant $D_2$
for a two-layer system $D_2$ as a multiplier. 
Thus, within the Mills model (\ref{energyMills}) 
the lability field of the AF phase has 
the same value for arbitrary values of $N$ 
\cite{Keffer73,Micheletti97,Zhedanov98,Dantas99}.
It coincides with that for a \textit{bulk} antiferromagnet
$H_{\mathrm{AF}}= \sqrt{K\,(2J+K)}$  (\ref{critfields2b}).

Note that this simple result for the stability limit of the AF phase 
holds only for Mills model because of its high symmetry.
For general models (see Eq.~(\ref{matrixAF})),
$H_{\mathrm{AF}}$ is an involved combination of the materials
parameters and depends on $N$.
For example, for the modified Mills model one derives
\begin{eqnarray}
D_N &  = & 
\left( H^4+ p_1 H^2 +q_1 \right) D_{N-4} \\
& & + 
\left( H^4+ p_2 H^2 +q_2 \right) \sum_{i=3}^{N/2}(-1)^{i}D_{N-2i}\,, \nonumber
\label{matrixAF}
\end{eqnarray}
where 
$p_1 = 3J^2 -(J+K)^2 - (J_s + K_s)^2$, $p_2 = p_1 - J^2$,
$q_1 = q_0 - J^2(J_s + K_s)^2$, $q_2 = q_0 - J^4$, and
$q_0 = [J^2- (J+K)(J_s+k_s)]^2$.
The determinants $D_{N-2j}$ ($j >2$) in the right part of 
Eq.~(\ref{matrixAF}) are sub-determinants and do not include ``surface'' terms.

\begin{figure}
\includegraphics[width=8.5cm]{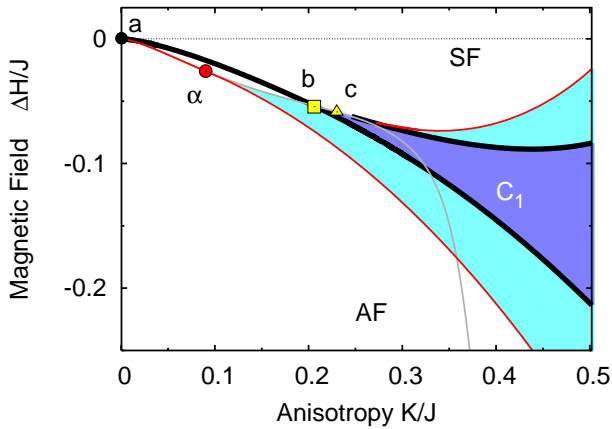}
\caption{
\label{PhDiagN4n6lowK}
(Color online)
Low anisotropy region of the
phase diagram for Mills model with $N$ = 6.
The magnetic field is given relative to the stability
limit of the AF phase by $\Delta H= H-H_{\mathrm{AF}}(K)$.
The canted phase C$_1$ is stable 
in the darker (blue) area.
It is metastable in the brighter (turquoise) areas.
Thick black lines give the first-order 
spin-flop transition from AF to SF and AF to C,
respectively. 
For anisotropies below point $\alpha$ 
only a first-order spin-flop from AF to the symmetric
SF-phase exists. At higher anisotropies above 
the point $b$, the first-order 
transition is from SF to the asymmetric canted C-phase.
For $ K_{\alpha} < K < K_b$, the canted phase 
exists only as a metastable state.
For $K > K_c$ above the tricritical point $c$, 
the transition from the asymmetric phase C$_1$ 
to the spin-flop phase is first-order.
Line $a$-$\alpha$-$b$ 
is the lower stability limit of the AF-phase,
along the line $\alpha$-$b$-$c$ 
the continuous transition between C and SF takes place.
Lines starting at point $\alpha$
are the metastability limits of the canted phase C.
}
\end{figure}

In Fig.~\ref{PhDiagN4n6lowK} the low-anisotropy
region is shown by $\Delta H$--$K$-phase diagrams 
for $N$=6 as representative for the general behaviour.
Here, magnetic fields are given relative to $H_{\mathrm{AF}}$, i.e.
$\Delta H \equiv H-H_{\mathrm{AF}}$. 
The stability limit for the antiferromagnetic
phase is always above the lower stability limits of the
symmetric and inhomogeneous SF-phase. 

Comparing the equilibrium energies in the AF and SF phase
we determine the field for the first-order transition
between these two phases (Fig.~\ref{PhDiagN4n6lowK}).
In the low-anisotropy limit this first-order transition line
and the two lability fields for the stability limits of the AF and SF phase,
respectively, 
are close to the value $H_0=\sqrt{2\,JH\,}$
from Eq.~(\ref{energy2c}). 
The difference between them
defining the co-existence region for metastable states
is of order $H_0(K/J) \ll H_0$.
\begin{figure}
\includegraphics[width=8.5cm]{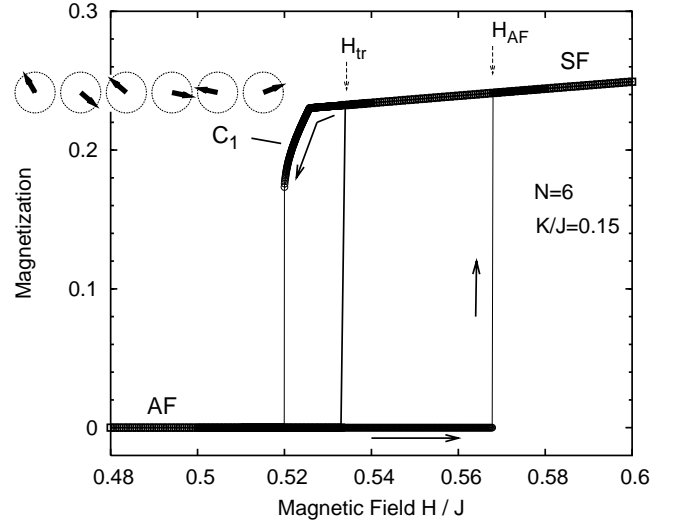}
\caption{
\label{MvsHN6}
 Hysteresis for Mills model $N=6$ and an anisotropy value,
where a canting instability of the spin-flop phase 
occurs in the metastability region $K_{\alpha} < K < K_b$. 
A resulting magnetic configuration of 
the canted C$_1$-state is shown.
Field $H_{\mathrm{tr}}$ for the first-order transition 
between antiferromagnetic and spin-flop-phase, 
and upper stability limit $H_{\mathrm{AF}}$ 
of the antiferromagnetic phase are indicated.
}
\end{figure}

\begin{figure}
\includegraphics[width=8.8cm]{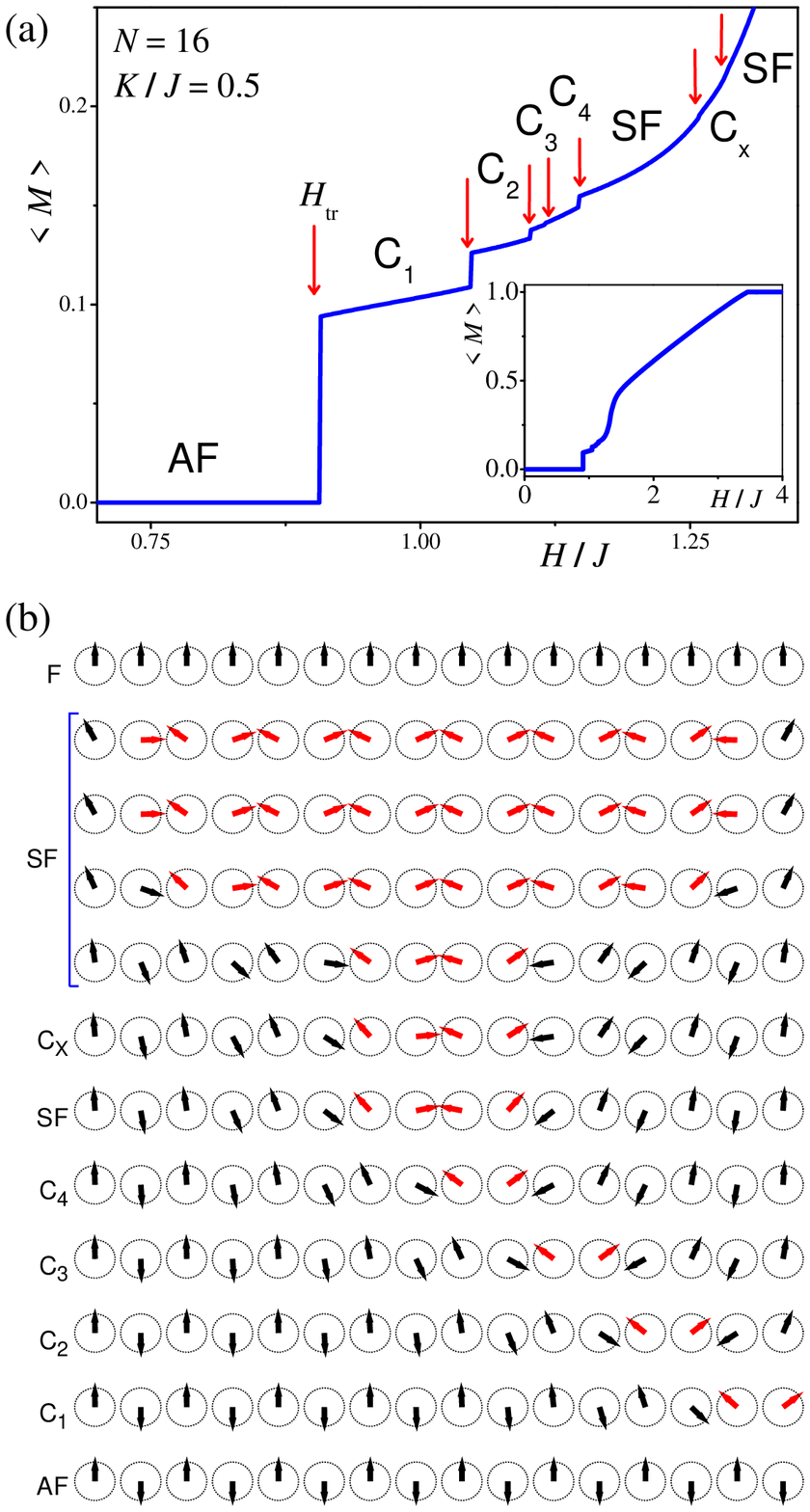}
\caption{
\label{ClocksN16}
(Color online)
Equilibrium solutions for Mills model 
with $N$=16 and intermediate 
anisotropy $K/J=0.5$.
In increasing field a series of transition
lead from antiferromagnetic to ferromagnetic
phase via several canted (C$_i$)
and (reentrant) spin-flop states.
Left panel: magnetization of the
equilibrium states(first-order transitions
are marked by arrows).
Right panel:
spin-configurations in the canted in
spin-flop states.
}
\end{figure}

Near the spin-flip transition from the 
SF to the F phase,
the deviations of $\mathbf{s}_i$ from 
the field directions are small
($\theta_i \ll 1$), and for model Eq.~(\ref{energy2}) 
the stability matrix in Eq.~(\ref{distortedEnergy}) 
becomes $\Phi_N(\tilde{\theta}_i)= 
\sum_{i,j=1}^{N}A_{ij}\theta_i\theta_j$
with $A_{ij}$ having a tridiagonal band matrix form
where nonzero elements occur only
in the main diagonal and the first side diagonals. 
In particular, for Mills model
$A_{ii} = H + K -J $,
$A_{i,\,i-1} = A_{i-1,\,i} =J$, and
the spin-flip or exchange field is a linear function of $K$
\begin{eqnarray}
H_{\mathrm{E}}^{(N)}(K) = 2J+K_f - K, \quad H_{\mathrm{E}}^{(N)}(0)= 2J+K_f\,,
\label{spinflip}
\end{eqnarray}
where $K_f$ is defined as the value of uniaxial anisotropy 
in the point $f$ where, depending on $N$, 
the line $H_{\mathrm{E}}^{(N)}(K)$ intersects 
line $H_{\mathrm{FM}} = 2\,J$.
The line $H_{\mathrm{FM}}$ is 
the transition line between the ferromagnetic
and ferrimagnetic phases.
The value $K_f$  can be derived analytically
as solutions of the equation $D_N(A_{ij}) =0$
with $A_{ii} = K_f -J $,
$A_{i,\, i-1} = A_{i-1,\, i} =J$ ( Table I);
$H_{\mathrm{E}}^{(N)}(0)= 2J+K_f $ is the spin-flip
field for zero anisotropy (in point $g$, Fig.~\ref{PDN6}).
In Ref.~[\onlinecite{PRB04a}] the spin-flip fields 
have been calculated for 
generalized isotropic models including biquadratic exchange.

For systems with larger anisotropies, 
an asymmetric canted phase C$_1$ occurs first 
as a \textit{metastable} state for $K_{\alpha} < K < K_{b}$,
which can be reached by 
a continuous canting of the spin-flop 
phase.  
A corresponding hysteresis 
around the first-order transition 
between AF and SF-phase 
is shown in Fig.~(\ref{MvsHN6}) with 
an example of the magnetic configuration C$_1$.
This canted phase C$_1$ is derived 
from elastically distorting
the collinear ferrimagnetic state FM$_1$ 
with a ferromagnetically aligned pair at the surface.
For even larger anisotropy $K_{b} < K$, the C$_1$ state
becomes a stable phase of the system, which is reached
from the AF-state through a first-order transition.
In the $H$ vs. $K$-phase diagrams, 
the magnetic fields for the upper and lower 
stability limit of the canted phase C$_1$ meet at
the critical point $\alpha$ at ($K_{\alpha}$, $H_{\alpha}$) 
This point also delimits the line for the 
canting instability of the \textit{metastable} SF-state. 
The critical line for the canting instability 
of the \textit{stable} SF-state ends 
in the critical point ($K_{b}$, $H_{b}$) on the 
line of first-order transitions $H_{\mathrm{tr}}$ between 
either the AF-phase and the SF-phase
below $K_b$, or the AF-phase and the C$_1$-phase above $K_b$.
This point $b$ located at ($K_{b}$, $H_{b}$) designates 
the lower anisotropy limit, where the phase C$_1$ and 
any asymmetric canted phase is stable for Mills models.
The topology of the phase diagrams 
in Fig.~\ref{PhDiagN4n6lowK} for the corner of 
low anisotropy $K < K_{\alpha}$ and fields 
describes the general behaviour for arbitrary $N$.
From our previous analysis, we have seen that 
no canted asymmetric phase may occur at low anisotropies.
The first canting instability at higher anisotropy 
will occur into a phase similar to the C$_1$ phase
with a flopped configuration at the surface.
For Mills models with various $N$ we have 
numerically determined the low-anisotropy parts of 
the phase diagrams and verified this general topology.
Coordinates of the two critical points $\alpha$ and $b$
for the canting instabilities 
are collected in Table~\ref{TabCritPointsMillN4b16}.
from numerical investigations of Mills models 
with $N=4,6,..,16$.

Magnetization curves corresponding to the equilibrium states, 
where the canted state C$_1$ is a stable phase 
are shown in Fig.~\ref{MvsHN6}.
For anisotropy $K > K_b$ further transitions 
and critical points occur depending on $N$. 
E.g., the transition between the C$_1$-phase 
and the spin-flop for $N=6$ becomes first-order 
above a tricritical point $c$.

\begin{table*}
\caption{
\label{TabCritPointsMillN4b16}
List of main critical points 
in the field-anisotropy 
phase diagrams for Mills model antiferromagnetic multilayers
with $N=2,4,\dots,16$. Field and anisotropy
values are given in units of $J$.
}
\begin{ruledtabular}
\begin{tabular}{lccclc}
$N$ & ($K_{\alpha}$, $H_{\alpha}$) & ($K_b$, $H_b$) & ($K_{\beta}$, $H_{\beta}$)\footnotemark[1]  & $K_e$\footnotemark[1] & $K_f$\\
\hline
4 & (0.160, 0.522)  & (0.300, 0.730) & (0.622102, 1.57956) & 0.847759     & 2$^{1/2}$ \\
6 & (0.090, 0.408)  & (0.206, 0.620) & (0.637223, 1.51922) & 0.842236     & 3$^{1/2}$ \\
8 & (0.051, 0.312)  & (0.120, 0.481) & (0.639260, 1.50798) & 0.842001     & $\sqrt{2+2^{1/2}}$ \\
10 & (0.034, 0.256) & (0.080, 0.394) & (0.639621, 1.50545) & 0.8419914    & $\sqrt{10+20^{1/2}}/2$ \\
12 & (0.024, 0.217) & (0.056, 0.332) & (0.639689, 1.50486) & 0.841990990  & $(6^{1/2}+2^{1/2})/2$ \\
14 & (0.019, 0.193) & (0.042, 0.286) & (0.639702, 1.50472) & 0.8419909729 & 1.9498\footnotemark[2] \\
16 & (0.014, 0.169) & (0.032, 0.251) & (0.639705, 1.50469) & 0.8419909721 & $\sqrt{2+\sqrt{2+2^{1/2}}}$ \\
\end{tabular}
\end{ruledtabular}
\footnotetext[1]{calculated from analytic expression with arbitrary precision.}
\footnotetext[2]{Numerical value given instead of a long analytic expression.}
\end{table*}

\subsection{Metamagnetism of strongly anisotropic systems}
At high enough uniaxial anisotropy 
only collinear phases are stable. 
For the infinite anisotropy limit, one can 
describe the model as an antiferromagnetic 
chain of classical Ising-spins.
In this limit, all collinear states coexist 
and transitions between them are first-order.
The equilibrium states and their transitions 
are found from the comparison 
of their different Zeeman and exchange energy. 
In Mills model (\ref{energyMills}) 
only two first order transitions take place. 
For increasing fields these are a transition 
from the antiferromagnetic (AF) state 
to a set of degenerate ferrimagnetic (FM) phases 
at $H=J$,
and the transitions from these FM-phase into 
the saturated phase (F) at $H=2J$.

Due to the high symmetry of Mills model,
it displays a remarkable degeneracy of the FM phase.
This degeneracy 
has important consequences for the 
structure of the phase diagram at finite anisotropy. 
Let us denote a ferromagnetic pair 
with configuration $\uparrow\uparrow$.
The two different antiferromagnetic domains
are (AF1)=$\uparrow\downarrow$ 
and with reversed spins (AF2)=$\downarrow\uparrow$.
The ferrimagnetic configuration with 
a flipped spin at the edge can be written 
(AF1)$^{N/2-1}$(F), where exponents 
denote the number of repetitions for a pair.
It is easy to see that all configurations 
of type FM$_n$=(AF1)$^{N/2-n-1}$(F)(AF2)$^{n}$ 
with $n=1,\dots,N/2$ have the same energy 
for Mills model (Fig.~\ref{Phases}~(b) panel (f)-(h)).
There are no further ferrimagnetic equilibrium
phases for this model.
For generalized models with differing
magnetic properties of individual layers 
the degeneracy of the FM phases will be lifted.
Then, the two transitions for Mills model
in the limit of infinite anisotropy 
will be replaced by sequences of metamagnetic 
transitions between various asymmetric collinear states.
The exact sequence will be subject to the set
of materials parameters for the individual layers.

Towards finite anisotropy, the collinear
phases will undergo characteristic instabilities
were the competition of Zeeman energy, exchange
and anisotropy will cause elastic distortions of 
these configurations. 
The stability limits for these
collinear phases can be calculated from 
the analytic expressions for zero
eigenvalues of their stability matrices $\mathbf{A}$ 
in Eq.~(\ref{distortedEnergy}).
These analytic expressions are derived similarly 
to the expressions for the upper stability limit 
$H_{\mathrm{AF}}$ of the AF phase, Eqs.~(\ref{critfields2b}), (\ref{matrixAF2}),
and for the lower stability limit or exchange field $H_{\mathrm{E}}^{(N)}$ 
of the F phase, Eq.~(\ref{spinflip}).
In principle, they can be evaluated with arbitrary precision.
But, the stability limits of the different $FM$ phases 
depend not only on $N$ but also on 
the particular realization FM$_l$,
i.e., on the location of the F-pair in the chain. 
For the ferrimagnetic phase of Mills model with $N=4$ 
there is only one lability line $H_{\mathrm{FM}} (K)$.
It can be written in the parametric
form 
\begin{eqnarray}
H_{\mathrm{FM}} =J \left(t - K(t)\right), \quad 
K=\frac{Jt(t^3-2t^2 + 2)}{2(t^3 - 2t -1)}\,.
\label{labilityFM}
\end{eqnarray}
The line $H_{\mathrm{FM}} (K)$ (\ref{labilityFM}) consists of
two branches meeting  in the point ($K^*$, $H^*$) where
$K^*=K(t^*)=0.62210$, $H^*=t^*-K(t^*)=1.57956$,
$(t^*)^2=2+2^{1/3}+4^{1/3}$.

\begin{figure}
\includegraphics[width=8.5cm]{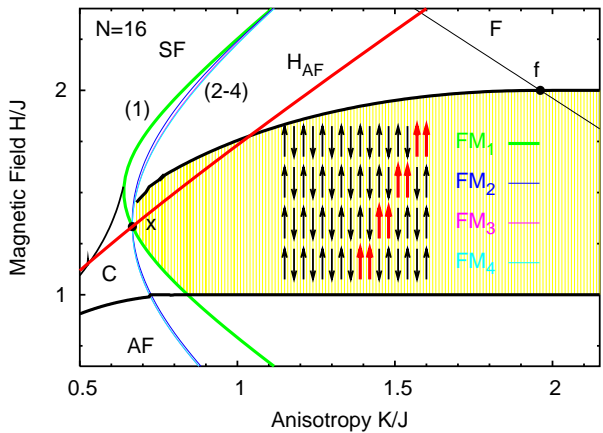}
\caption{
\label{FM1to4N16}
(Color online)
High-anisotropy phase diagram for Mills model and $N$=16.
The degenerate ferrimagnetic phases are stable 
in the shaded area.
The collinear phases FM$_1$ and FM$_2$ to FM$_4$
are limited by the lability lines (1) and (2-4).
The latter are nearly degenerate and not resolved.
}
\end{figure}

For arbitrary $N$, the stability limits 
for the various energetically 
degenerate ferrimagnetic phases FM$_l$ are different.
Here, $l=1 \dots [(N-2)/2]$ and  square brackets $[x]$ 
denote the largest integer $l \leq x$.
However, they display a certain systematics for Mills models. 
This can be understood from the weakened exchange stiffness
at the surfaces, i.e. the surface cut, 
which distinguishes the state FM$_1$ with an F-pair
at the surface from all other realizations F$_l$ with $l>1$
(see Fig.~\ref{PDN6}(b) for the simplest case $N=6$).
The generic behaviour of these lines 
is demonstrated for the case $N=16$ in Fig.~\ref{FM1to4N16}.
The lower branch for FM$_1$ occurs always at higher 
fields and anisotropies. 
This is the expected consequence of the surface cut.
The FM$_1$ structure is more easily 
distorted and plays a special
role in the intermediate anisotropy range. 
In all phase diagrams, at a certain 
section of this line a continuous 
transition between the FM$_1$ phase and 
the corresponding C$_1$-phase occurs 
(see Fig.~\ref{PDN6}(b)).
This transition line at low fields and high anisotropies, 
starts at the critical end-point ($K_e$, $H_e$) 
on the first-order line between the AF and FM-phase. 
In the phase diagrams for $N=6$, 
this section ends at a multicritical point $x$.
For general $N$, the other 
end of the line of continuous transitions 
between the FM$_1$ and the C$_1$ phase 
depends on $N$ because at 
higher fields other canted phases C$_l$ with $l>1$ 
may intervene. 
Thus, the series of co-existing collinear states FM$_l$ 
at high anisotropy gives rise to series of corresponding canted phases C$_l$ by
elastic distortions. These canted phases, however, are stable 
in {\em different} regions of the phase diagram towards intermediate anisotropies.
Thus, the series of canted phases in increasing fields  
starts with C$_1$ and, for the highly symmetric Mills model, 
it follows the sequence $l=1,\dots\,$  (Fig.~\ref{ClocksN16}(b)).

For the first canting transition between the ferrimagnetic state FM$_1$ 
and the corresponding canted phase C$_1$, 
along the line for the stability limit of the FM$_1$ phase,
there is a minimum value of anisotropy $K_{\beta}$.
The corresponding critical point $\beta$ coincides with the 
upper limit of fields $H_{\beta}$ for the 
metastability limit of the canted phase C$_1$. 
The values of these two characteristic points 
are listed in Table~\ref{TabCritPointsMillN4b16} 
for Mills models $N=$4 to 16.
There is only a weak dependence on $N$ 
for the coordinates for the critical points $\beta$ and $e$.
This means, that there is also only a weak shift of the 
region of stable and metastable FM-states.
Further, the minimum anisotropy values for the stability 
region of the other collinear phases FM$_l$ with $l>1$ 
are always larger than $K_{\beta}$.
Thus, canted phases may occur in the region 
rather well circumscribed by the area shown for the
simplest phase diagrams for $N=4$ and 6.
Interestingly, all the stability lines for these
collinear phases FM$_l$ $l=1,\dots\;$ 
cross at the point $x$ at $(K_x,\,H_x)=(4/3J,\, 2/3J)$ 
for Mills model and arbitrary $N$. 
One can show this by a similar recursion for the
eigenvalues of their tridiagonal stability matrices
as used by Dantas et al. to calculate $H_{\mathrm{AF}}$.\cite{Dantas99} 
This point is also visited by the line $H_{\mathrm{AF}}(K)$. 
For all systems with an even-odd number of layers $N$=6, 10, etc., 
the point $x$ is a special multicritical point,
where the collinear analogue of the symmetric spin-flop phase
FM$_{[(N-2)/2]}$ becomes degenerate with 
the inhomogeneous symmetric SF- state. 
Thus, the first-order transition line between 
the corresponding canted phase C$_{[(N-2)/2]}$ 
and the SF-phase ends in $x$ as in the phase
diagram for $N=6$ shown in Fig.~\ref{PDN6}.
This further degeneracy of Mills model 
yields some stability and simplicity 
of the general features of its phase diagrams.

As a caveat, we finally have to mention additional low-symmetry phases
which cannot be foreseen from the considerations on the stable 
collinear phases AF, FM$_l$, and F and their elastic distortions in the phase diagram.
Such an intermediate phase C$_x$ appears already in the phase diagram of 
Mills model for $N=6$ in the region between the FM$_l$ phases and
the region of the stable SF-phase. 
The C$_x$ phase can be derived from an 
elastic distortion of 
the collinear phases $\uparrow\downarrow\uparrow\uparrow\uparrow\uparrow$, 
which do not arise as stable states.
Similar low-symmetry canted phases also appear 
in phase diagrams of Mills model for $N>6$.
For generalized models, the region of stability of
these canted low-symmetry phases will strongly 
depend on details of the materials parameters of type Eq.~(\ref{energy2}). 
Further energy terms and/or disorder 
in magnetic parameters of the layers 
can stabilize further phases 
in the intermediate region of the phase diagram. 
In the high-anisotropy 
limit of such generalized models, cascades of metamagnetic
transitions between ferrimagnetic collinear phases exist.
From these states various canted phases can be derived 
in the range of intermediate anisotropies 
$K_i/J_i \stackrel{<}{\sim} 1$.
The competition between all these phases will lead 
to very complicated magnetic phase diagram 

\section{Reorientation in weakly anisotropic multilayers }\label{WeakSection}

\subsection{Surface and volume interactions}\label{SurfandVolSub}
In this section, we re-analyze the magnetic states
of an antiferromagnetic multilayer stack, 
and their evolution in an applied field,
by analytical methods.
In the limit of weak anisotropy, 
the micromagnetic energy (\ref{energy1}) 
can be represented by a system of interacting dimers 
and by a continuum form.
Thus, we consider the model (\ref{energy1}) 
for weak anisotropies, $K_i,\,K'_i \ll J_i$.
Further, we only study systems with collinear antiferromagnetic
(zero-field) ground state. 
Thus, the strengths of 
the biquadratic exchange constants
is limited to the range, $0< \widetilde{J}_i <J_i/2 $.\cite{Demokritov98}
First, we rewrite the general energy Eq.~(\ref{energy1}) in this limit. 
The resulting expression allows to recognize
the main effects expected in this limit without explicit calculations. 
We group the moments of the superlattice into pairs 
as in a two-sublattice antiferromagnetic system.
Starting from the first layer we combine the 
$N$ moments along the antiferromagnetic chain
into $N/2$ pairs
$(\mathbf{s}_{2j-1}, \mathbf{s}_{2j})$ with $j=1,..,N/2$.
For each of these pairs, we introduce the vectors of 
net magnetization ${\mathbf{M}}_j$ and the staggered vector $\mathbf{L}_j$
\begin{eqnarray}
\label{mlj}
\mathbf{s}_{2j-1} = {\mathbf{M}}_j + \mathbf{L}_j, \quad
\mathbf{s}_{2j} = {\mathbf{M}}_j - \mathbf{L}_j \,.
\end{eqnarray}
These transformations are similar to those
applied for the two-layer system, Eqs.~(\ref{energy2a}) and (\ref{energy2b}).
From $|\mathbf{s}_i| = 1$ follows 
that  ${M}_j^2+L^2_j = 1$  
and ${\mathbf{M}}_j \cdot \mathbf{L}_j=0$.
The energy $\Theta_N$ of Eq.~(\ref{energy1}) 
can be rewritten as a function
of ${M}_j$ and the angles $\phi_j$ between $\mathbf{n}$
and unity vectors $\mathbf{l}_j=\mathbf{L}_j/|\mathbf{L}_j|$ 
and expanded  with respect to the small parameters ${M}_j \ll 1$. 
Omitting constant and higher order terms, one derives 
\begin{eqnarray}
\label{energy3}
\widetilde{\Theta}_N & = & 
\sum_{j=1}^{N/2}\Lambda_j {M}_j^2 
-2\sum_{j=1}^{N/2}\left(\mathbf{H} \cdot {\mathbf{M}}_j \right)
- \sum_{j=1}^{N/2} \widetilde{K}_i \cos^2 \phi_j \nonumber \\
& \, & +\frac{1}{2} \sum_{j=1}^{N/2-1}J_{2j} (\phi_{j+1}-\phi_j)^2
+\Xi({M}_j, \phi_j)\,, 
\end{eqnarray}
where %
$\widetilde{K}_j=(K_{2j-1} + K_{2j})/2-K'_{2j-1}-(K'_{2j-2}+K'_{2j})/2$,
$\Lambda_j = 2(J_{2j-1}-2\widetilde{J}_{2j-1})
             +(J_{2j-2}-2\widetilde{J}_{2j-2})
             +(J_{2j}-2\widetilde{J}_{2j})$
for $j=2,\dots,N/2-1$, and
$\widetilde{K}_1= (K_1 + K_2)/2-K'_1-K'_2/2$,
$\widetilde{K}_{N/2}=(K_{N-1}+K_N)/2-K'_{N-1}-K'_{N-2}/2$;
$\Lambda_1 = 2(J_1-2\widetilde{J}_1)+(J_2-2\widetilde{J}_2)$,
$\Lambda_{N/2} = 2(J_{N-1}-2\widetilde{J}_{N-1})
                  +(J_{N-2}-2\widetilde{J}_{N-2})$.
Finally, the last expression in Eq.~(\ref{energy3})
collects terms that are linear with respect 
to $(\phi_{j+1}-\phi_j)$,
\begin{eqnarray}
\label{energyXi}
\Xi ({M}_j, \phi_j)  &  = & \\ 
 & -& \sum_{j=1}^{N/2-1}(J_{2j}-\widetilde{J}_{2j})
 ({M}_j +{M}_{j+1})
(\phi_{j+1}-\phi_j)\,. \nonumber
\end{eqnarray}
An independent minimization with respect to ${\mathbf{M}}_j$
(see details in Ref.~[\onlinecite{FNT86, PRB02}]) yields 
\begin{eqnarray}
\label{mj}
{\mathbf{M}}_j=\Lambda_j^{-1} 
\left[ \mathbf{H} - \left(\mathbf{H} \cdot \mathbf{l}_j \right)
\mathbf{l}_j \right]\,.
\end{eqnarray}
It follows directly from (\ref{mj}) that
${M}_j = H \Lambda_j^{-1} \sin (\phi_j - \psi)$,
where $\psi$ is the angle between the field $\mathbf{H}$
and the easy axis $\mathbf{n}$ in the model of Eq.~(\ref{energy1}).

The independent minimization with respect to ${\mathbf{M}}_j$ 
is justified because the exchange interactions dominate 
the energy and pairs of neighbouring moments do not deviate strongly 
from antiparallel orientation in the limit of weak anisotropy and fields.
This establishes the relations Eq.~(\ref{mj}) 
between the components of the net magnetization and 
the orientation of the staggered vector. 
In other words Eq.~(\ref{mj}) fixedly connects 
the net magnetizations, as auxiliary degrees of freedom,
to the vectors $\mathbf{l}_j$ and the applied field.
This approach reduces 
the chain of $N$ magnetic moments $\mathbf{s}_i$ 
into an equivalent system of $N/2$ 
unity vectors $\mathbf{l}_j$. 
Each site of this chain corresponds 
to a two-sublattice antiferromagnet 
or a \textit{dimer}.
Substituting (\ref{mj}) into Eq.~(\ref{energy3})
we obtain the following expression for the
energy of these \textit{interacting dimers} 
\begin{eqnarray}
\label{PhiJ}
\widetilde{\Theta}_N   & = &  
-\sum_{j=1}^{N/2} \bar{\Phi}_j\cos2(\phi_j-\bar{\phi}_j)   \\
& & +
\frac{1}{2} \sum_{j=1}^{N/2-1}J_{2j} (\phi_{j+1}-\phi_j)^2
+\Xi( \phi_j)\,, \nonumber
\end{eqnarray}
\begin{eqnarray}
\label{PhiJbar}
\bar{\Phi}_j  =
\frac{1}{2 \Lambda_j}
\sqrt{\left( H^2 \cos2\psi -H_j^{2} \right)^2
+ H^4 \sin^2 2\psi\;}\,,
\end{eqnarray}
\begin{eqnarray}
\label{phijbar}
 \tan 2 \bar{\phi}_j & = &   H^2 \sin 2\psi/(H^2 \cos2\psi -H_j^{2}), 
\end{eqnarray}
\begin{eqnarray}
\label{Hj}
 H_j & = &  \sqrt{\widetilde{K}_j \Lambda_j }\,, 
\end{eqnarray}
where
\begin{eqnarray}
\label{energyXi1}
\Xi (\phi_j) & = & 
-H\sum_{j=1}^{N/2-1}\Omega_j \sin(\phi_j - \psi)
(\phi_{j+1}-\phi_j), \\
\label{Omegaj}
\Omega_j  & = &  J_{2j} \left(\Lambda_{j}^{-1}+\Lambda_{j+1}^{-1} \right)\,.
\end{eqnarray}
The minimization with respect to ${\mathbf{M}}_j$ according to Eq.~(\ref{mj}) 
and the representation of the energy by the form (\ref{PhiJ}) 
generalizes simplified dimerization transformation 
that have been considered in Refs.~[\onlinecite{pss04,JAC}].

The energy of the interacting dimers Eq.~(\ref{PhiJ})
includes first the sum of  their ``self''-energies, 
then an exchange stiffness energy given by 
the  term quadratic with respect to differences $(\phi_{j+1}-\phi_j)$,
and a specific energy contribution $\Xi(\phi_j)$, 
defined in Eq.~(\ref{energyXi1}). 
The terms in $\Xi(\phi_j)$
arise due to the variation of 
the magnetic parameters along the chain.
The energy $\Xi(\phi_j)$ can be written 
in the form of a ``Zeeman energy'' 
for the staggered magnetization vectors 
\begin{eqnarray}
\label{energyXi2}
\Xi( \phi_j) & = & 
-\left[ \Omega_1(\mathbf{H} \cdot \mathbf{l}_1)
-\Omega_{N/2-1}(\mathbf{H} \cdot \mathbf{l}_{N/2}) \right] \\
& &-\sum_{j=1}^{N/2-2} \left( \Omega_{j+1} - \Omega_j \right)
\left(\mathbf{H} \cdot \mathbf{l}_{j+1} \right)\,. \nonumber
\end{eqnarray}
The dimensionless coefficients $\Omega_j$ 
are ratios of 
exchange constants defined in Eq.~(\ref{Omegaj}).
The first two terms in Eq. (\ref{energyXi2}) involve 
the endmost dimers, i.e., they have the character of 
a \textit{surface} energy, which is imposed by the exchange cut.
The sum in (\ref{energyXi2}) describes similar ``internal''  contributions 
arising due to any variation of the exchange couplings along 
the antiferromagnetic chain. 
This energy contribution disappears in models with equal exchange
interactions in internal layers as in the regular 
and modified  Mills models Eqs.~\ref{energyMills}) and (\ref{MillsModified}),
respectively.

\subsection{Physical mechanism of the ``surface spin-flop'' phenomena}
The magnetic energy of the low-anisotropy antiferromagnetic multilayers
in the form of interacting dimers  (\ref{PhiJ})
elucidates the competing forces responsible for
the field-driven reorientation processes.
Let us compare energy (\ref{PhiJ}) with that of an isolated dimer.
For a localized pair of $\mathbf{s}_{2j-1}$ 
and $\mathbf{s}_{2j}$ spins (i.e. $j$-th dimer)  
a minimization via Eq.~(\ref{mj}) yields
\begin{eqnarray}
\label{Phi2}
& & \Theta_2^{(j)}  =-\Phi_j^{(0)} \cos2(\phi_j-\bar{\phi}_j^{(0)})\,,
\end{eqnarray}
with an amplitude factor 
\begin{eqnarray}
\label{Phi2amplt}
& & 
\Phi_j^{(0)}= \frac{1}{4J} \sqrt{\left( H^2 \cos2\psi -\bar{H}_j^{2} \right)^2
+ H^4 \sin^2 2\psi\;}\,,
\end{eqnarray}
and a ``phase''
\begin{eqnarray}
\label{Phi2phase}
& & 
\tan 2 \bar{\phi}_j^{(0)}  =  H^2 \sin 2 \psi/(H^2 \cos 2 \psi -\bar{H}_j^{2}),
\end{eqnarray}
where 
$\bar{H}_j^{2} = [(J_{2j-1}-2\widetilde{J}_{2j-1})(K_{2j-1}+K_{2j}+2K'_{2j-1})]^{1/2}$.
The energy in the form (\ref{Phi2}) coincides
with that of a two-sublattice antiferromagnet
and is a generalization of the model Eq.~(\ref{energy2b}).
The Eq.~(\ref{Phi2phase}) for the phase $\bar{\phi}_j^{(0)}$
is known as \textit{N{\'e}el equation}.\cite{Neel36}
It determines the equilibrium states of the antiferromagnet
$ \phi= \bar{\phi}_j^{(0)} + \pi k$ ($k=0,1,\dots$). 
The amplitude $\Phi_j^{(0)}$ from Eq.~(\ref{Phi2amplt}) 
equals the potential
barrier between the wells at $ \phi= \bar{\phi}_1 + \pi k$.
A magnetic field applied along the easy axis reduces the potential barrier. 
When the field reaches the threshold field  $\bar{H}_j$ 
it causes the spin-flop transition.
For dimers incorporated into the interacting chain the parameters
of the self-energies are modified due to the exchange
coupling and additional anisotropy contributions as seen by comparing
Eqs.~(\ref{PhiJ})-(\ref{phijbar}) with Eqs.~(\ref{Phi2})-(\ref{Phi2phase}).
Therefore, within the chain the dimers have \textit{different} threshold fields
{\em and} they are elastically coupled with neighbouring pairs.
Due to the couplings the flopping of the individual dimers
in their individual threshold fields $H_j$ Eq.~(\ref{PhiJ}) are hampered.
Instead the chain only can transform into the flopped phase
when the flopped configurations are energetically advantageous
throughout the whole system.
Generally the differences of the dimer self-energies along the chain
causes spatial modulations of any noncollinear magnetic states.
The inhomogeneous spin-flop and canted phases
in Mills model (Fig. \ref{Phases}) exemplify such spin-configurations.

However, the energy contribution $\Xi( \phi_j)$ Eq.~(\ref{energyXi}) provides
{\em another mechanism} of magnetic-field-induced reorientation
imposed {\em by the variation of the exchange interactions} 
along the antiferromagnetic superlattice, in particular 
by the exchange cut at its ends.
This mechanism is due to 
the influence of the linear energy terms (\ref{energyXi1}),
which favour a rotation of the staggered vector.
As can be seen from the equivalent Eq.~(\ref{energyXi2}),
an instability of the collinear configuration
is caused by the ``Zeeman terms'' that are linear 
in the staggered vectors $\mathbf{l}_j$.
Generally, the first term related to the two surfaces 
will dominate, and this difference will 
favour a rotation of these $\mathbf{l}_j$.
This enforces an inhomogeneous spin-flop phase 
above a certain field.
As was shown in the previous section
in strong anisotropy systems the exchange
cut leads  to flips of the magnetization
and a transition into collinear FM phases, 
which are also inhomogeneous.
In low-anisotropy systems, 
under the dominating influence of the exchange interactions,
the influence of this ``local'' defect is spread 
along the chain and stabilizes a spatially inhomogeneous structure. 

Thus, we have the following important conclusion.
There are two different mechanisms of the field-induced
reorientation in finite antiferromagnetic superlattices:
(i) One of them is connected with a switching of 
the potential wells in the energy of 
the uniaxial antiferromagnetic units.
This mechanims is similar to the usual field-driven 
spin reorientation in (low-anisotropy) bulk antiferromagnets 
and two-layer systems.
Therefore, it is a common \textit{spin-flop} mechanism.
(ii) The other mechanism is due to variation of the exchange
coupling along the superlattice and, in particular,
the exchange cut at the end of the stack.
This type of reorientation transition can only exist
in finite antiferromagnetic superlattices and has 
no analogue in bulk antiferromagnetism. 

The interplay of these two mechanisms rules
the formation and evolution of the magnetic states 
in the system.
Depending  on the values of the material parameters
different types of magnetization processes
can be realized in the general model (\ref{energy1}).
In the low-anisotropy systems, owing to the dominance
of exchange interactions, it is the second effect due 
to the cut exchange at the surfaces that dominates the 
field-driven reorientation transition. 

As important cases for applications, 
we consider in more detail 
the highly symmetric Mills models Eqs.~(\ref{energyMills}) 
and (\ref{MillsModified}).
Both models are composed of identical internal layers. 
For the modified Mills model Eq.~(\ref{MillsModified})
$\Lambda_1=\Lambda_{N/2} = 2(2J_s +J)/(1+\zeta_s)$,
and $\Lambda_j= 4J$ for $j=2,\dots,N/2-1$. 
For the regular Mills model Eq.~(\ref{energyMills})
$\Lambda_j= 4J$ for $j=1,\dots,N/2$. 
The energy (\ref{PhiJ}) with $\psi=0$ reduces to
\begin{eqnarray}
\label{energyPhiMills}
\tau_N & = &
\sum_{j=2}^{N/2-1}\frac{\left(H^2-H_{\mathrm{B}}^2 \right)}{8J}\cos2\phi_j \\
&& +\frac{J}{2} \sum_{j=1}^{N/2-1}(\phi_{j+1}-\phi_j)^2 + \Phi_s, \nonumber
\end{eqnarray}
where $H_{\mathrm{B}} = \sqrt{4JK}$.
The last contribution in (\ref{energyPhiMills}) is due to the 
finite length of the chain. 
It involves only the two last 
dimers at both ends of the chain, 
i.e., it represents the specific surface effects
for the finite antiferromagnetic stack.
For the regular Mills model the isolated dimer energy (\ref{Phi2}) 
reduces to the form of Eq.~(\ref{energy2c}), 
and the surface energy becomes
\begin{eqnarray}
\label{energySurfaceM}
\Phi_s &=&
\frac{1}{6J}\left(H^2-H_S^2 \right)
\left(\cos2\phi_1 +\cos2\phi_{N/2} \right) \nonumber\\
& & -\frac{7}{12} H \left(\cos \phi_1 -\cos \phi_{N/2}  \right)
\nonumber\\
& & +\frac{1}{12} H \left(\cos\phi_2 -\cos \phi_{N/2-1}  \right)\,,  \\
 H_S &  =  & \sqrt{3JK}\,. 
\end{eqnarray}

In the case of the modified Mills model, the contribution $\Phi_s$
has the modified form
\begin{eqnarray}
\label{energySurface}
\widetilde{\Phi}_s &=&
\frac{(1+\zeta_s)^2}{8(2J_s+J)}\left(H^2-\widetilde{H}_S^2 \right)
\left(\cos2\phi_1 +\cos2\phi_{N/2} \right) \nonumber\\
 & &  -  U_1 H \left(\cos \phi_1 -\cos \phi_{N/2}  \right) \nonumber\\
& & +U_2 H \left(\cos\phi_2 -\cos \phi_{N/2-1}  \right)\,.  
\end{eqnarray}
The threshold field $\widetilde{H}_S$ 
and the coefficients, $U_1$ and $U_2$,
are defined by
\begin{eqnarray}
\label{Hs}
 \widetilde{H}_S  & = &  
\frac{\sqrt{2(K_s+K)(2J_s+J)}}{(1+\zeta_s)}, \\
 U_1 & = & \frac{J(6 \zeta_s-1)+2J_s(4\zeta_s -3)}{4(2J_s+J)}, \\
U_2 & = & \frac{J(2\zeta_s +1) -2J_s}{4(2J_s+J)}\,. 
\end{eqnarray}
The threshold field for the endmost dimers is $H_S =\sqrt{3/2} H_0 $
and for internal dimers $H_{\mathrm{B}} =\sqrt{2} H_0$.
Because the spins in the chains have additional exchange couplings,
these thresholds are larger than the threshold field $H_0$
for an isolated pair.
This reinforcing by exchange stiffness 
for the bound dimers increases the values
of the threshold fields for the coupled chain.

\subsection{Multilayers with large N. Continuum model}

The limit of multilayers with large $N$
and the limit of infinite antiferromagnetic systems is
best discussed going over to a continuum description.
For the regular Mills model (\ref{energyMills}) 
with arbitrary $N$
the transition into the flopped state
occurs closely to $H_{\mathrm{tr}} \approx H_0$ 
given by Eq.~(\ref{energy2c}),
sufficiently below the dimer threshold fields 
$H_S$ and $H_{\mathrm{B}}$
(Fig.~\ref{BSFN100}).
This means that this transition is imposed by the exchange cut.
These results can be easily understood from 
the solutions for the spin-configurations in the flopped state
Eqs.~(\ref{SFsol}), (\ref{sol8N}), and (\ref{sol10N}).
The magnetization vectors for all internal layers
can be combined into pairs with antiparallel
magnetizations. 
The system effectively behaves as an isolated dimer 
consisting only of the endmost spins with energy (\ref{energy2}).
Correspondingly the flopping field equals
the threshold field $H_0$ of this isolated pair.
This result is common for
systems with arbitrary values of $N$.

\begin{figure}
\includegraphics[width=8.5cm]{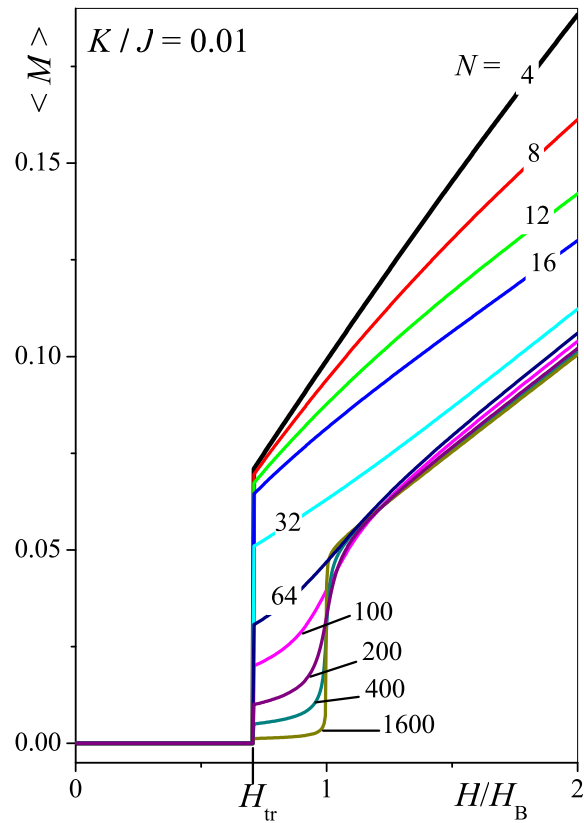}
\caption{
\label{BSF1600}
(Color online)
Magnetization curves for Mills model 
with low anisotropy $K/J$=0.01 
and large N
in the vicinity of the ``bulk spin-flop''.
}
\end{figure}

\begin{figure}
\centerline{\includegraphics[width=9.1cm]{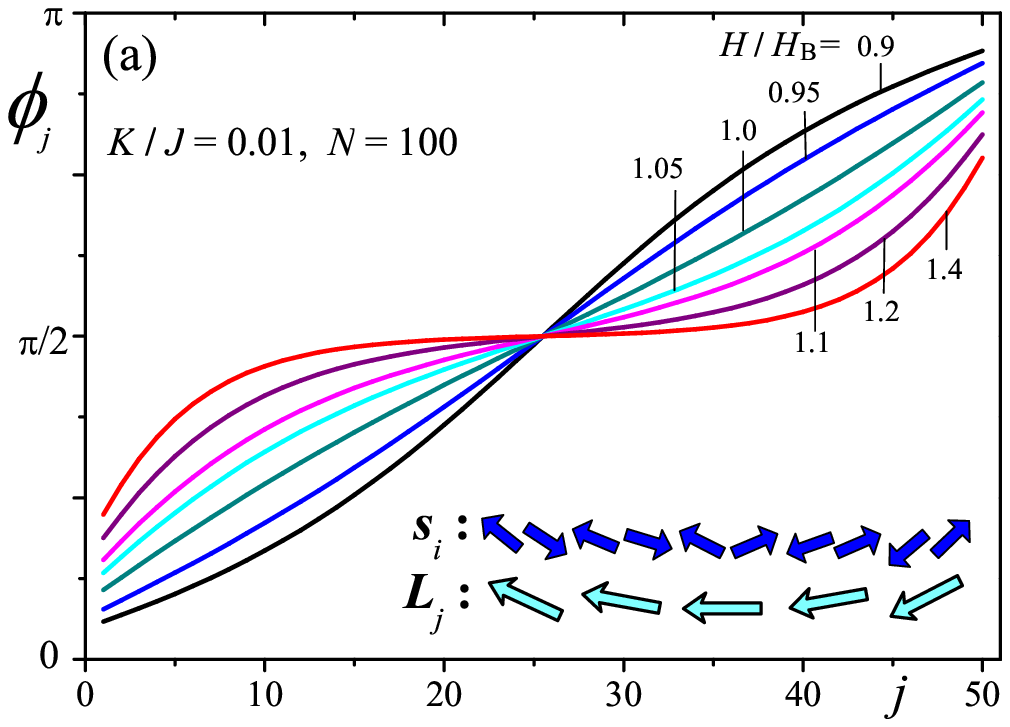}  }
\centerline{ \includegraphics[width=8.5cm]{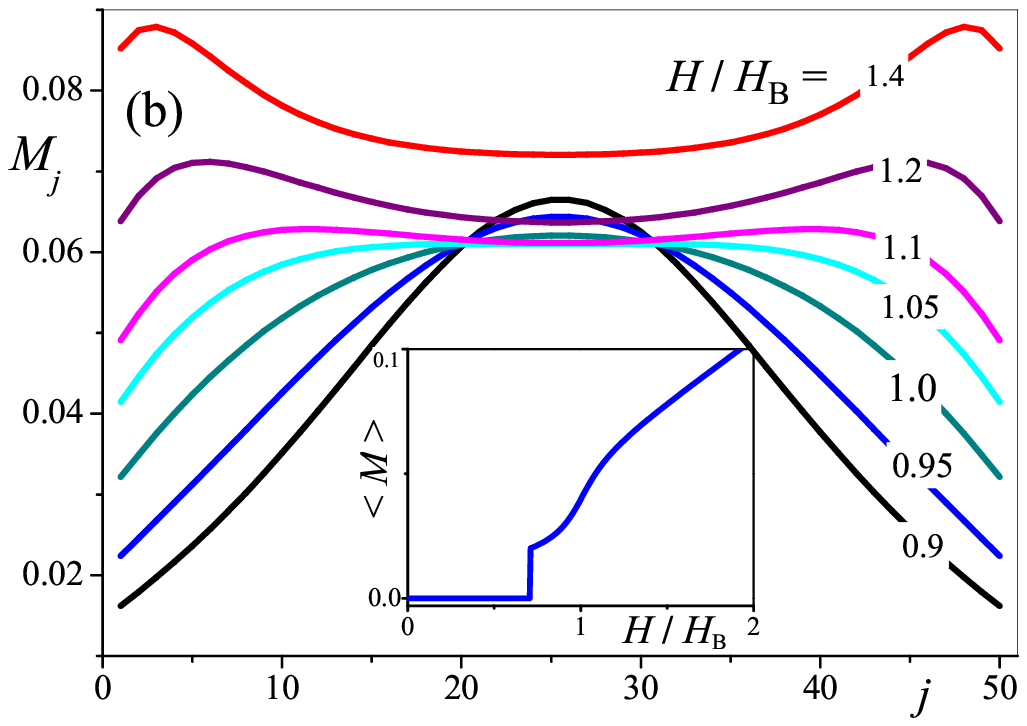} }
\caption{
\label{BSFN100}
(Color online)
Evolution of configurations for Mills model 
with low anisotropy $K/J$=0.01 and $N=100$ 
in the vicinity of the ``bulk spin-flop'':
(a) local net magnetic moments $M_j$
(b) orientation of local antiferromagnetic vectors (according to the Eq.~(\ref{mlj}).
Inset shows average magnetization $m$ vs. field.
}
\end{figure}

However, above the critical field $H_0$
the evolution of the system
remarkably changes with increasing $N$
(Fig.~\ref{BSF1600}).
The multilayer with low $N$ are characterized
by a large magnetization jump at the transition field 
and a nearly linear increase 
of the magnetization up to the flip field.
With increasing $N$ the magnetization jump
at $H_0$ gradually decreases. 
Concurrently, a steep section of the magnetization 
curve is found around fields 
with values close to $H_{\mathrm{B}}$ (see Eq.~(\ref{energyPhiMills})).
Finally, for $N \gg 1$ the magnetization curves 
develop a strong kink around this value. 

The magnetic-field driven transformation of the dimer energies
in Eq.~(\ref{energyPhiMills}) explains this 
peculiar behaviour.
In the transition field $H_{\mathrm{tr}} \approx H_0 $ 
the dimer self-energy terms 
in (\ref{energyPhiMills}) still favour
the antiferromagnetic mode $(\mathbf{l}_j || \mathbf{n})$.
The threshold fields are exceeded at higher fields
($H > H_S$  for endmost and $H > H_{\mathrm{B}}$ for internal dimers).
In superlattices with few layers the endmost 
and internal dimers give comparable contributions
to the magnetic energy. 
The difference in their internal energies
suppresses drastic reorientation effects
at the threshold fields $H_S$  and $H_{\mathrm{B}}$.
With increasing number of layers the relative
energy contribution of the internal dimers 
for the total energy (\ref{energy3})
gradually increases. 
Then, the magnetic energy of 
the internal layers plays the dominant role in the formation 
of the flopped configurations.
Thus, the endmost dimers does not hamper
the reorientation effects in the threshold
field $H_{\mathrm{B}}$.
Below the threshold field,  $H < H_{\mathrm{B}}$,
the antiferromagnetic phase with $\mathbf{l}|| \mathbf{n}$
corresponds to the minima of the internal dimers
and the inhomogeneous spin-flop phase consists
of two antiferromagnetic domains with
antiparallel orientation of the staggered vectors
(Fig.~\ref{BSFN100}).
These two regions are separated by a 180-degree domain wall 
with flopped spin configurations in the center
$(\mathbf{l}\bot \mathbf{n})$ 

For  $H > H_{\mathrm{B}}$  the potential wells for the internal dimers
switch into $(\mathbf{l}\bot \mathbf{n})$ configuration.
Around the field $H_{\mathrm{B}}$ the center of the domain wall gradually
extends and sweeps out the regions with antiferromagnetic
configuration towards the surfaces of the stack.
This drastic transformation between the two
configurations within most of the bulk of 
the antiferromagnetic stack 
causes a prominent anomaly of the magnetization
curves near the field $H_{\mathrm{B}}$ (Fig.~\ref{BSF1600}).
Above $H_{\mathrm{B}}$ the net magnetization $M_j$ develops 
two symmetric maxima close to the surfaces, which
may be observable in experiment.

Asymptotically with $N\rightarrow\infty$,
the magnetization curve  approaches that of
the usual spin-flop in a bulk uniaxial antiferromagnet.
But, this reorientation occurs within
{\em the same magnetic phase}, and no phase
transition is connected with this process.
Rather for any finite value of $N$ 
the phase transition still occurs 
between the antiferromagnetic and inhomogeneous
spin-flop phase in the critical field $H_{\mathrm{tr}}$ 
as a first-order process.
As was mentioned above for large $N$ this
field-induced phase has the character of 
a domain wall between two antiferromagnetic states.
Non-collinear states arise in the 
central region of the stack, where the small
total magnetization of the configuration is
concentrated.
For fixed small anisotropy the transition 
is accompanied by a jump of the magnetization
at the transition field $H_{\mathrm{tr}}$. 
The magnitude of this jump decreases with the numbers
of layers $N$ (Fig.~\ref{BSF1600}).
Hence, at first glance we have a paradox phase diagram
for Mills model:
a drastic field-driven change of the magnetization
is not related to a phase transition, 
while a \textit{real} phase transition is noticeable
only by a small jump of the magnetization,
that vanishes for large $N$.
However, this has a clear physical foundation
because the transition at $H_{tr}$ is related to the 
surface effect and its visible effects should vanish 
for $N\rightarrow\infty$, whereas the crossing-over
towards the flopped configuration in the ``bulk'' 
of the multilayer stack should approach a true 
spin-flop transition for $N=\infty$.

A transition into the inhomogeneous spin-flop state
means that the free boundaries 
cause an inhomogeneity far in the interior of
the finite system.
Close to the boundaries 
the magnetic configuration resembles 
that of the two antiferromagnetic collinear domains. 
This structure is consistent with the 
properties of semi-infinite antiferromagnetic chains
described by Micheletti et al. \cite{Micheletti97}
In the phase diagrams for these systems 
(even in the large anisotropy limit)
a highly degenerate phase occurs,
where a localized inhomogeneous configuration 
is situated at arbitrary distance
from the surface.\cite{Micheletti97}
For finite antiferromagnetic chains with weak anisotropy, 
the mutual influence of both surfaces will determine
a unique state with 180~degree wall-like configuration 
in the center.
Generally, such a symmetric configuration 
will be found for antiferromagnetic layers, 
when the core of this configuration is wide enough
to interact with both surfaces.
For the finite systems, 
the simple structure of the phase diagram, showing only 
a SF phase with solutions preserving 
mirror symmetry about the center of the layer
in intermediate fields between the AF phase
and the saturated F phase,
is restricted to low anisotropy systems.
At sizeable anisotropy, the asymmetric canting 
complicates the phase-diagram.

The energy of the modified Mills model Eq.~(\ref{MillsModified})
provides a simple way to introduce a continuum form
of energy (\ref{energyPhiMills}).  
For $N \gg 1$ and low anisotropy $K\ll 1$
the energy (\ref{energy3}) can be converted to
\begin{eqnarray}
\label{energyCont}
\Delta W & = &  W - W_{\mathrm{AF}}  =  \nonumber \\
& &\int_{-d/2}^{d/2}\left[
A \left(\frac{d\phi}{d\xi}\right)^2 +
\left(\frac{H^2 - H_{\mathrm{B}}^2}{16 A} \right)
 \sin^2  \phi \right]d\xi \nonumber \\
 && + W_s|_{\xi= \pm d/2}\,
\end{eqnarray}
with the exchange constant $A = J a/2$.
The multilayer thickness is $d = Na$ 
with $a$ the ``periodicity'' length of the multilayer.
The zero of energy scale for $\Delta W$
is shifted to the energy $W_{\mathrm{AF}}$
of the antiferromagnetic state ($\phi = 0$). 
The last term is the  surface energy given by
\begin{eqnarray}
\label{energyContS}
W_{s} & = & \frac{(1+\zeta_s)^2}{4(2J_s+J)}\left(H^2-\widetilde{H}_S^2 \right)
\sin^2 \phi(\xi)  \nonumber\\
 & - &\mathrm{sgn}(\xi) \,\left(\zeta_s -\frac{1}{2}\right) H 
\left[\,\cos \phi(\xi) - 1 \,\right] 
\end{eqnarray}
Eq.~(\ref{energyCont}) describes the energy
of a plate of thickness $d$ for 
a bulk easy-axis antiferromagnet
with the spin-flop field $H_{\mathrm{B}}$.
It is the continuum counterpart of
the discretized model Eq.~(\ref{energyPhiMills}).
The surface energy $W_s$ (\ref{energyContS})
includes a common antiferromagnetic contribution
(the first term) and a specific Zeeman energy 
imposed by the exchange cut.
Due to mirror symmetry of the inhomogeneous 
spin-flop phase the boundary conditions are 
$\phi_1 \equiv \phi(\xi= - d/2)$
and $\phi_2 \equiv \phi(\xi= d/2) = \pi - \phi_1$.
By solving the Euler equation for
the energy functional (\ref{energyCont})
with the boundary condition
$2A(d\phi/d\xi)|_{\xi= \pm d/2}=-\partial W_s(\phi1)/\partial \phi$
one obtains a set of parametrized profiles
$\phi(\xi, \phi_1)$.
The further optimization of the energy
with respect to the parameter $\phi_1$
yields the equilibrium distribution of the
staggered vector within the multilayer of finite thickness. 
These solutions can generally be written as elliptic functions.

In the limit of infinite thickness,
$N\rightarrow\infty$ the boundary
values of $\phi$ correspond to 
the configurations in antiparallel domains
of the antiferromagnetic phase,
$\phi_1 (-\infty)$ =0, 
$\phi_2 (\infty) =\pi$.
In this case the variational
problem for the functional
(\ref{energyCont})
is equivalent to that
of an isolated magnetic wall 
with ``effective uniaxial anisotropy'' 
$K^* =(H^2 - 2H_0^2)/(16A)$.
The corresponding
analytical results for the wall
parameters have been derived by Landau and Lifshitz
\cite{Landau35}.
For Mills model this solution 
has been analysed in \cite{Mills68}.

This limiting case of the infinite
chain provides a simple physical
explanation of the phase transition
into the inhomogeneous spin-flop phase.
According to Eqs. (\ref{energyCont}),
(\ref{energyContS})
the flop of the surface
staggered vector $\phi(\infty)=0$
into antiparallel position
$\phi(\infty) =\pi $ 
yields a gain of surface energy
$\Delta W_s = -(2\zeta_s -1)H$.
By this process
a domain wall is generated 
which requires a positive energy contribution
$4\sqrt{AK^*} = \sqrt{H_{\mathrm{B}}^2 -H^2\,}$.
The balance between these competing energy contributions 
is reached at the critical field
\begin{equation}
\label{balance}
H_{\mathrm{tr}} = H_0/\sqrt{2\zeta_s^2 -2 \zeta_s +1\,}\,.
\end{equation}

In particular, for the regular Mills model
$\zeta_s =1$ and the transition field in Eq.~(\ref{balance}) equals $H_0$.
This simple energetics allows to formulate
a clear thermodynamic reason for the transition into the inhomogeneous 
spin-flop phase provided by the exchange cut.
\textit{
The anti-aligned magnetization vector
at the non-compensated surface is overturned and 
reduces the Zeeman energy
at the expense of the formation of a planar defect
in  the center of the superlattice}, which has
the character of a domain wall.

Because the energy gain in the inhomogeneous spin-flop phase,
$\Delta W_s (\zeta_s)= -(2\zeta_s -1)H$,
is proportional to the non-compensated 
magnetization of the surface layer, 
this transition into the flopped state
strongly depends on the \textit{net magnetization} of the endmost layers.
Partial compensation of the surface
magnetization is tantamount 
to a reduction of the parameter $\zeta_s$.
This reduction decreases $\Delta W_s (\zeta_s)$,
and increases the critical field $H_{\mathrm{tr}}$ (\ref{balance}).
For $\zeta_s = 1/2$
the energy gain $\Delta W_s (\zeta_s)$ equals zero
and $H_{\mathrm{tr}}$ reaches
the threshold field $H_{\mathrm{B}}$.

We come to an important conclusion.
The exchange cut provides the stabilization
of the flopped phase only under condition
of sufficiently strong surface magnetization.
In the regular Mills model the magnetizations of the layers are
assumed to have fixed values.
The properties of the endmost layers $j=1$ and $N$
are described by the same integral phenomenological parameters 
as the layers $j=2,\dots,N-1$ in the ``bulk''. 
Only the exchange cut reflects the confinement of 
this antiferromagnetic system.
In the continuum limit (\ref{energyCont}) 
the surface cut is represented by
the surface contributions in the energy 
that describe the effective coupling of 
the staggered vector to the applied field.
The existence of surfaces with net 
non-compensated magnetization is justified 
for antiferromagnetic superlattices with in-plane magnetization. 
Strong intra-layer exchange interactions and
weak stray field effects 
favour single domain states of the individual
ferromagnetic layers in these multilayer stack 
and at the surfaces.
In other systems, various mechanisms can 
cause reductions of the non-compensated
magnetization at the surfaces, 
such as crystallographic and magnetic imperfections,
formation of antiferromagnetic multidomain states etc.
A reduction of the total magnetization in the surface layers
can strongly reduce the non-compensated magnetization
and suppresses eventually the formation of the flopped states.
For the continuum model (\ref{energyCont}) this occurs
for $\zeta_s < 1/2$.

The surface energy (\ref{energyContS}) 
also includes the first term that 
has the conventional form of 
a (local) antiferromagnetic unit 
with a modified threshold field $\widetilde{H}_S$
from Eq.~(\ref{Hs}). 
According to many experimental observations, 
the magnetic parameters $J_s$, $K_s$ and $\zeta_s$ 
can be strongly modified by surface-induced interactions, 
see, e.g., Refs.~[\onlinecite{Stiles99,Johnson96}].
Correspondingly $\widetilde{H}_S$ in (\ref{energyContS})
can vary in a broad range.
Generally, considering models with modified magnetic surface
properties, the volume energy (\ref{energyCont}) 
and the surface energy (\ref{energyContS}) 
may favour different magnetic configurations 
in certain intervals of applied magnetic field.
This competition can stabilize inhomogeneous phases 
with continuous rotation of the magnetization vectors
along the thickness. 
The occurrence of such \textit{twisted} states 
under pinning (or anchoring) influence of 
the surfaces is a rather general effect 
in orientable media.
In particular, they are known 
to occur in various classes of liquid crystals 
as the so-called \textit{Freedericksz} effect
\cite{Freed33,deGennes} and in
ferromagnetic materials.\cite{Goto65,Nolting00,Berkowitz99}
Spiraling in exchange spring magnets and exchange bias 
systems also belongs to this class of phenomena.\cite{Berkowitz99}
The phenomenological theory of such states
in antiferromagnetic nanolayers has
been developed in Ref.~[\onlinecite{PRB03}].
It was shown that non-collinear twisted phases
can arise as solutions for magnetic states 
under anchoring-effects at the surfaces.
In contrast to inhomogeneous spin-flop 
states stabilized by the exchange cut 
the twisted phases arise due 
to pinning or distortive effects 
of surface-induced interactions on the magnetic states.
Future analysis of generalized Mills models
should concentrate on the combined effects 
of these surface interactions.

\section{Discussion}\label{Discussion}
\subsection{Solving the ``surface spin-flop'' puzzle}\label{Puzzle}

The exchange cut is the primary driving force that causes 
the specific reorientation effects
in antiferromagnetically ordered multilayers
and stabilizes the unique magnetic states unknown
in other classes of magnetic systems.
The pioneering studies by Mills 
and co-workers \cite{Mills68,Wang94}
have introduced the notion of 
a surface-induced instability 
in confined antiferromagnets \cite{Mills68} 
and of the novel reorientation effects 
in antiferromagnetic superlattices \cite{Wang94}.
Mills formulated the basic idea 
that, in a confined antiferromagnet,
uncompensated surface magnetization 
causes the instability of the collinear state
in the applied magnetic field 
quite below the common (bulk) spin-flop.
This transition from the antiferromagnetic 
to a ``surface spin-flop'' state should result 
in flopping a few layers of spins near the surface,
i.e., they would turn by nearly 90 degree.\cite{Mills68}
This picture was 
improved and detailed by Keffer and Chow,\cite{Keffer73}, 
and supplemented by results of numerical simulations.
\cite{Wang94,Papan98,Rakhmanova98,Mills99}
It constitutes the recent scenario of a ``surface spin-flop''.
According to this picture the flopped states 
are nucleated initially at the surface 
and in increasing fields
this surface state moves 
into the depth of the sample as an antiferromagnetic domain wall:
\cite{Keffer73,Rakhmanova98}
\textit{``When the external field exceeds the surface spin-flop
field, the surface moment, initially antiparallel to the
field, rotates nearly by 180$^{\circ}$. 
In effect, a twist is applied to
one end of the structure. 
A domain wall is then set up, in
an off center position in the finite structure. 
[$\dots$]
With further
increase in field, the wall undergoes a series of discontinuous
jumps, as it migrates to the center of the structure.
[$\dots$]
The domain wall becomes centered in the structure, 
and then with further increase in field broadens, 
to open up as a flower to evolve into a bulk 
spin-flop state. The angle between
the spins and the external field is less 
at and near the surface than in the center 
of the structure.``} \cite{Mills99}

The detailed investigations of Mills model
(\ref{energyMills}) necessitate
a considerable revision of the surface spin-flop scenario expanded 
in \cite{Mills68,Wang94} and some other papers
\cite{Keffer73,Rakhmanova98,Mills99}.
This scenario of the surface spin-flop
confuses three different types
of the solutions for Mills model 
(\ref{energyMills}), see the phase diagrams 
in Figs.~\ref{Phases} and \ref{PDN6}.
The ``flopping of a few layers of 
spins near the surface'' is inspired by
solutions for ferrimagnetic and canted phases
in systems with sizeable anisotropy.
The picture of the domain wall movement 
into the center of the superlattice ``in a sequence
of discrete hops'' \cite{Rakhmanova98}  is related
to cascades of phase transitions between
canted phases in superlattices with moderate
anisotropy (Figs.~\ref{ClocksN16}).
Finally, the flower-like broadening of the centered 
domain wall poetizes the evolution of
the inhomogeneous spin-flop phases in
fields higher than $H_{\mathrm{B}}$ (Fig.~\ref{BSF1600}).

Thus, the common scenario of the surface spin-flop
combines elements that belong to 
\textit{different} solutions 
for \textit{different} values of the
control  parameters ($K/J$, $H/J$, $N$)
of the model (\ref{energyMills}). 
In Ref. \cite{PRB04}, it was shown that 
the  equations minimizing energy (\ref{energyMills}), 
as well as general models (\ref{energy1}),
do not include solutions for 
surface-confined (\textit{localized}) states, 
which were assumed to occur 
at a ``surface spin-flop" transition''. 
These models own only
well-defined ``volume'' phases 
and transitions between them.

The term ``surface spin-flop''
designates the reorientation
into the inhomogeneous spin-flop phase
at $H_{\mathrm{tr}}$ (line $a-b$ in Fig.~\ref{PDN6}).
Therefore, it is a double misnomer.
This transition does not take place at the \textit{surface}
because it involves the reorientation 
of all spins along the superlattice, i.e., it 
has a ``volume'' character. 
And it is not a proper \textit{spin-flop} because 
it is induced by the exchange cut
rather than a switching of the potential
wells as in the common spin-flops in
bulk antiferromagnets.

\subsection{Notes on the experimental observations 
of surface spin-flop phenomena}\label{Experiments}

The concept of a ``surface spin-flop'' 
is commonly  applied to analyse experimental results
in antiferromagnetic superlattices 
\cite{Wang94,Felcher02,Lauter02,Nagy02}.
However, the application of an erroneous concept is dangerous. 
In particular, quantitative conclusions 
about magnetic materials parameters 
from the observed reorientation transitions 
can lead to wrong results.
The \textit{lability field of the antiferromagnetic states} 
$H_{\mathrm{AF}}$ plays the prime role 
in the common ``surface spin-flop'' scenario.
Because the surface spin-flop 
was believed to arise as a local surface
instability of the collinear phase exactly 
at the critical field $H_{\mathrm{AF}}$,
this was considered as a transition field 
into the surface spin-flop state \cite{Mills68,Keffer73}.
In reality a (\textit{volume}) first-order 
transition between the antiferromagnetic and 
inhomogeneous spin-flop phases occurs at $H_{\mathrm{tr}}$ 
(e.g., the line $a$-$b$ in Fig. \ref{PDN6}),
which is lower than  $H_{\mathrm{AF}}$ (line a$\lambda$) 
and larger than another lability field $H_{\mathrm{SF}}$ 
(line $a$-$\alpha$).  
The interval $H_{\mathrm{SF}} < H < H_{\mathrm{AF}}$ is a
metastability region of these competing 
phases (Fig. \ref{PhDiagN4n6lowK}).
In the low-anisotropy limit the metastablity region
is extremely small and  these characteristic field are
all close to the value $H_0$ from Eq.~(\ref{energy3}).
In the limit of large anisotropy
the lability field $H_{\mathrm{AF}}$ is much larger
than the transition field between AF and
FM phases (Fig.~\ref{PDN6}).

\textit{The ``bulk'' spin-flop field} 
is also considered as important element of the common scenario.
Starting at $H_{\mathrm{AF}}$ the expansion of 
the surface spin-flop phase is completed 
in fields exactly equal to the value
of the spin-flop transition in a ``bulk'' antiferromagnet
having the same values of the magnetic parameters
as in model (\ref{energyMills}).
For low-anisotropy systems this field equals $H_{\mathrm{B}}$ 
and is $\sqrt{2}$ times larger than the ``surface spin-flop'' $H_0$.
This field corresponds to the threshold field for the internal dimers 
as given in Eq.~(\ref{energyPhiMills}).
In systems with large numbers of layers $N$,
a strong reorientation of the flopped states
occurs in the vicinity of this field.
No phase transition is connected with
this process, however, it is marked by a noticeable
anomaly of the magnetization curve (Fig. \ref{BSF1600}).
Thus, the  magnetization curve anomalies are connected 
with the transition into the flopped state 
at $H_{\mathrm{tr}} \approx H_0$ and with 
a smooth reorientation near $H_{\mathrm{B}}$ 
that does not involve a real transition.
The ratio $H_{\mathrm{B}}/H_{\mathrm{tr}}$ is about $\sqrt{2}$.
A similar anomaly within the spin-flop state is also observed
in systems with rather large anisotropy, where the
spin-flop phase is preceded by one or several canted phases.
However, a glance at the phase diagram, e.g., Fig.~\ref{PDN6} 
shows that there is no simple quantitative relation between the various 
reorientation anomalies observable in such multilayer
systems with sizeable anisotropy.

Magnetic-field-induced reorientation 
transitions were investigated
in high-quality Fe/Cr(211)
antiferromagnetic superlattices
\cite{Wang94,Felcher02}.
In Ref.~[\onlinecite{Wang94}] magnetization curves 
for Fe/Cr (211) superlattices with
strong uniaxial anisotropy were
measured by a SQUID magnetometer and 
by longitudinal magneto-optic Kerr effect.
The magnetization curves for both
investigated multilayers 
with even number of layers
Cr(100)/[Fe(40)/Cr(11)]$_{22}$
and
Cr(100)/[Fe(20)/Cr(11)]$_{20}$
demonstrate
close correspondence to theoretical
results for Mills model.
According to \cite{Wang94}
the values of the antiferromagnetic
coupling between the layers
is $JM_0^2$ = 0.275 erg/cm$^2$
and uniaxial anisotropy is
$K M_0^2$ = 0.06 erg/cm$^2$,
The ratio $K/J$ = 0.22 shows
that these multilayers belong
to the systems at intermediate anisotropy in the 
phase diagram that display cascades of canted phases.
Indeed, the characteristic anomalies
in the field derivatives of the magnetization 
reveal a series of such reorientation transitions.
The asymmetric character of these
transitions is demonstrated by Kerr measurements
(see Fig.~3~(b) in Ref.~[\onlinecite{Wang94}]).
A cascade of canted phases
has also been observed in another Fe/Cr system \cite{Felcher02}.
In this paper a Cr(100)/[Fe(14)/Cr(11)]$_{20}$ system 
has been investigated
with $JM_0^2$ = 0.405 erg/cm$^2$
and
$K M_0^2$ = 0.06 erg/cm$^2$.
The ratio $K/J$ = 0.15 means
that this superlattice also evolves
in the applied field via a cascade
of canted phases.
It should be noted that the first-order transitions
between these different magnetic phases allows
for phase co-existence in rather wide field ranges 
(see Refs.~\onlinecite{PRB04a,JMMM04}]).
All these processes may involve multidomain states.
In the case of multilayers with effective in-plane anisotropy,
the stabilization of domain structures will be subject to 
imperfections. In particular, interface roughness will
lead to magnetic charges or leaking dipolar stray fields.
The corresponding domain structures is determined
by the defect structure of the multilayer and will have
an irregular appearance in general
(see, e.g., chap. 5.5.7 in Ref.~[\onlinecite{Hubert98}]).

In contrast, multilayers with perpendicular anisotropies 
constitute a novel class of artificial confined
antiferromagnets, where well-defined and regular domain structures
such as stripes or bubbles can be found.
These are layered systems 
as antiferromagnetically coupled multilayers [CoPt]/Ru \cite{Hellwig03}, 
or Fe-Au superlattices \cite{Zoldz04}.
These strongly anisotropic systems correspond
to the right side of the phase diagrams in 
Figs. \ref{Phases}, \ref{PDN6} and are characterized by a
number metamagnetic jumps \cite{Hellwig03,JMMM04}.
Due to strong demagnetization fields the magnetization
processes in these superlattices are accompanied
by a complex evolution of multidomain states
\cite{Hellwig03,Hellwig03b,JMMM04,Hellwig05}.
Artificial layered systems of this kind
with a controlled variation of 
distinct magnetic states in vertical direction
can be considered as artificial metamagnets \cite{JMMM04}.

\section{Conclusions}\label{Conclusions}
In this work, we have provided a complete solution
for the basic micromagnetic model 
of an antiferromagnetic superlattice 
with ideal non-compensated surfaces under a field along the easy axis.
We have shown how one can systematically enumerate 
and describe the magnetic phases and their transitions 
for such structures. 
The puzzle of the variable appearance of ``surface spin-flop'' phenomena 
has been resolved by the re-construction of the phase-diagrams 
and of the limiting cases for this model.
To this end various methods had to be introduced that can 
be used for generalized models.
Analytical tools can be efficiently used for all
collinear or highly symmetric phases, and for 
the case of weak anisotropies.
Extensions as given by the models Eqs.~(\ref{energy1}) --(\ref{MillsModified}),
that include further magnetic coupling terms, additional anisotropies etc.,
should be made the subject of further work. 
In particular, the question of 
competing surface-couplings and partially compensated surfaces in finite
antiferromagnetic stacks should be addressed. 
In such systems, a 
competition between a genuine inhomogeneous spin-flop phase and 
twisted states takes place. 

In systems with intermediate anisotropies comparable to the 
indirect interlayer exchange within antiferromagnetic superlattices,
one has to expect very complicated phase-diagrams. 
Still, such situations can be analysed by 
the micromagnetic methods developed here.
However, it is vital to use clear concepts of magnetic phase transitions 
and clean definitions that designate the driving forces 
behind the varieties of field-driven reorientation processes in 
confined antiferromagnets.
We emphasize that the notion of a ``surface spin-flop'' is erroneous
because the relevant magnetic energy terms that drive both the canting instabilities 
at surfaces and the transition into the inhomogeneous spin-flop phase are not related 
to a balance between effective anisotropies and Zeeman energy 
in these finite antiferromagnets. 
The transitions experienced by the type of finite antiferromagnets 
with non-compensated surfaces, as investigated here, 
are always related to the exchange cut.
Finally, for the artificial antiferromagnetic systems 
composed of mesoscale ferromagnetic units,
the first-order transitions are of crucial importance.
The phase-coexistence between states with finite
magnetization will give rise to stable domain 
structures and hysteretic behaviour 
in these systems owing to demagnetization effects.

\begin{acknowledgments}
This work was financially supported by DFG 
through SPP 1133, project RO 2238/6-1.
A.\ N.\ B.\ thanks H.\ Eschrig for support and
hospitality at the IFW Dresden. 
We acknowledge T.\ Laubrich 
for support with numerical calculations,
and we thank D. Elefant, O. Hellwig, 
J. Meersschaut, V.Neu, for discussion.
\end{acknowledgments}


\end{document}